\documentclass[a4paper,11pt]{article}

\usepackage{jheppub}
\usepackage{amsmath,amssymb,amscd}
\usepackage{graphicx}
\usepackage{xcolor}
\usepackage{url}
\usepackage{nccmath} 
\usepackage{bbm} 

\graphicspath{ {fig/} }

\makeatletter
\def\hlinewd#1{%
  \noalign{\ifnum0=`}\fi\hrule \@height #1 \futurelet
   \reserved@a\@xhline}
\makeatother

\makeatletter
\renewcommand\@fpheader{}
\renewcommand\@journal{}
\makeatother

\newcommand{\LabeledGraph}[2]{
\begin{minipage}{0.18\textwidth}
$\includegraphics[width=\textwidth]{#2}\hspace*{-16ex}\raisebox{9ex}{{#1}}\hfill$
\end{minipage}
\hspace*{2ex}
} 
\definecolor{darkgreen}{rgb}{0.,.3,0}
\definecolor{darkblue}{rgb}{0.0,0.0,0.5}
\newcommand{\ep}{\epsilon}
\newcommand{\mint}[1]{\scalebox{0.85}{#1}}
\newcommand{\pole}[1]{\textcolor{darkblue}{#1}}
\newcommand{\order}[1]{\textcolor{darkblue}{\mathcal{O}(#1)}}
\newcommand{\casimir}[1]{{#1}}
\newcommand{\ud}{\mathrm{d}}

\title{The Four-Loop $\mathbf{\mathcal{N} = 4}$ SYM Sudakov Form Factor}

\preprint{MSUHEP-21-029, P3H-21-082, TTP21-043}

\author[a]{Roman N. Lee,}
\author[b]{Andreas von Manteuffel,}
\author[b]{Robert M. Schabinger,}
\author[c,d]{\\Alexander V. Smirnov,}
\author[e,d]{Vladimir A. Smirnov,}
\author[\,f]{and Matthias Steinhauser}

\affiliation[a]{Budker Institute of Nuclear Physics, 630090 Novosibirsk, Russia}
\affiliation[b]{Department of
Physics and Astronomy, Michigan State University,\\
East Lansing, Michigan 48824, USA}
\affiliation[c]{Research Computing Center, Moscow State University,
119991, Moscow, Russia}
\affiliation[d]{Moscow Center for Fundamental and Applied Mathematics,
119992, Moscow, Russia}
\affiliation[e]{Skobeltsyn Institute of Nuclear Physics of Moscow State University,
119991, Moscow, Russia}
\affiliation[f]{Institut f{\"u}r
Theoretische Teilchenphysik,
Karlsruhe Institute of Technology (KIT),\\
76128 Karlsruhe, Germany}

\emailAdd{roman.n.lee@gmail.com}
\emailAdd{vmante@msu.edu}
\emailAdd{schabing@msu.edu}
\emailAdd{asmirnov80@gmail.com}
\emailAdd{smirnov@theory.sinp.msu.ru}
\emailAdd{matthias.steinhauser@kit.edu}

\abstract{
We present the Sudakov form factor in full color $\mathcal{N}=4$ supersymmetric Yang-Mills theory to four loop order and provide uniformly transcendental
results for the relevant master integrals through to weight eight.
}

\begin{document}
\unitlength1cm
\maketitle
\allowdisplaybreaks[1]

\section{Introduction} 
\label{sec:intro}

Over the past few decades, enormous progress has been made towards understanding the perturbative properties of $\mathcal{N}=4$ super Yang-Mills theory~\cite{Brink:1976bc,Gliozzi:1976qd} (referred to hereafter as $\mathcal{N}=4$ SYM).
The high symmetry of this theory allows one to gain structural insights and test computational approaches at higher orders in perturbation theory, possibly pioneering ideas with future applications to phenomenologically-important theories such as Quantum Chromodynamics; see \cite{Henn:2020omi} for a recent review.
Particularly impressive perturbative results have been obtained in the leading planar-color limit of the theory.
A celebrated example is the light-like cusp anomalous dimension~\cite{Korchemsky:1987wg}, which can be obtained to arbitrary loop order~\cite{Beisert:2006ez} in this limit.
At full color, analytical 4 loop results have been obtained for the cusp anomalous dimension \cite{Grozin:2017css,Henn:2019rmi,Lee:2019zop,Henn:2019swt,Huber:2019fxe}, the collinear anomalous dimension \cite{Dixon:2017nat,Agarwal:2021zft} and the universal anomalous dimension of twist-2 operators \cite{Kniehl:2021ysp}.
Planar on-shell scattering amplitudes are particularly well understood~\cite{Bern:2005iz,DelDuca:2010zg} and have been bootstrapped up to 7 loops for 6 legs~\cite{Caron-Huot:2019vjl}.
Form factors for local operators are more complex quantities, but have nevertheless been obtained up to 5 loops and 3 on-shell states~\cite{Dixon:2020bbt} in the planar limit; see also~\cite{Brandhuber:2018xzk,Ahmed:2019yjt,Sever:2020jjx,Lin:2020dyj} for further recent work on planar and non-planar color form factors and \cite{Penante:2016twu,Yang:2019vag} for reviews.

Sudakov form factors are among the most basic form factors, defined as the matrix elements of a length-two local operator between a two-particle state and the vacuum. They allow for a particularly transparent discussion of infrared poles in gauge theories~\cite{Mueller:1979ih,Collins:1980ih,Sen:1981sd,Magnea:1990zb,Sterman:2002qn,Ravindran:2004mb,Moch:2005id,Moch:2005tm,Dixon:2008gr,Falcioni:2019nxk}.
In particular, the $1/\ep^2$ and $1/\ep$ poles of the form factor are determined by the cusp and collinear anomalous dimensions, respectively.
In $\mathcal{N}=4$ SYM, the scalar Sudakov form factor was calculated to 2 loops in~\cite{vanNeerven:1985ja} and to 3 loops in~\cite{Gehrmann:2011xn}.
The integrands have been obtained at 4 loops in \cite{Boels:2012ew} and at 5 loops in \cite{Yang:2016ear}.
The poles of the 4-loop form factor have been presented in~\cite{Huber:2019fxe,Agarwal:2021zft}.
In this paper, we present the calculation of the full-color 4-loop Sudakov form factor through to the finite part.
Our calculation is based on techniques that we developed for the calculation of 4-loop form factors in Quantum Chromodynamics~\cite{Henn:2016men,Lee:2016ixa,Lee:2017mip,Lee:2019zop,vonManteuffel:2016xki,vonManteuffel:2019wbj,vonManteuffel:2019gpr,vonManteuffel:2020vjv,Agarwal:2021zft,Lee:2021uqq}.

The remainder of this paper is organized as follows.
In section~\ref{sec:conventions} we define the Sudakov form factor that we consider and review the known reduced integrand at 4 loops.
In section~\ref{sec:ffmasters} we describe our calculation of the relevant master integrals to transcendental weight 8.
In section~\ref{sec:results} we give the result for the Sudakov form factor.
In section~\ref{sec:conclusions} we conclude.

\section{Reduced integrand}
\label{sec:conventions}

The Sudakov form factor in $\mathcal{N}=4$ SYM that we consider in this paper is defined as
\begin{equation}
\label{eq:ffsudakov}
F = \frac{1}{N} \int\!\mathrm{d}^4x \, e^{-i q \cdot x} \,
  \langle \phi_{12}^a(p_1)\phi_{12}^b(p_2) \vert\,
  \left(\phi_{3 4}^c \phi_{34}^c\right)\!(x)
  \,\vert 0 \rangle ,
\end{equation}
where the expectation value of a local length-two operator is computed between the \mbox{vacuum} and a state with 2 on-shell scalar particles.
The Lorentz scalars $\phi^a_{12}$ carry subscripts corresponding to the $\mathbf{6}$ representation of the $R$-symmetry group $SU(4)_R$ and a superscript corresponding to the adjoint representation of the gauge group $SU(N_c)$.
The overall normalization $N$ is chosen such that the tree level contribution is normalized to 1.

For the kinematics, we have $p_1^2=p_2^2=0$ due to the massless on-shell states, such that the form factor depends only on the external scale $q^2=(p_1+p_2)^2$.
For the perturbative expansion of the form factor, we abbreviate
\begin{align}
a = \frac{N_c\, g^2}{16\pi^2}, \qquad\qquad
z = \frac{4\pi}{e^{\gamma_\text{E}}} \left(\frac{\mu^2}{-q^2-i0}\right),
\end{align}
where $g$ is the original bare coupling of the model,
$\gamma_\text{E} \approx 0.577216$ is Euler's constant,
$\epsilon=(4-d)/2$, $d$ is the number of space-time dimensions used to regularize the theory, and $\mu^2$ is the 't Hooft scale.
We define
\begin{equation}
\label{eq:ffpert}
F = \sum_{L=0}^\infty a^L z^{L\ep} F_L
\end{equation}
with $F_0=1$ and set $q^2=-1$ without loss of generality.

\begin{figure}
\centering
\LabeledGraph{(5)}{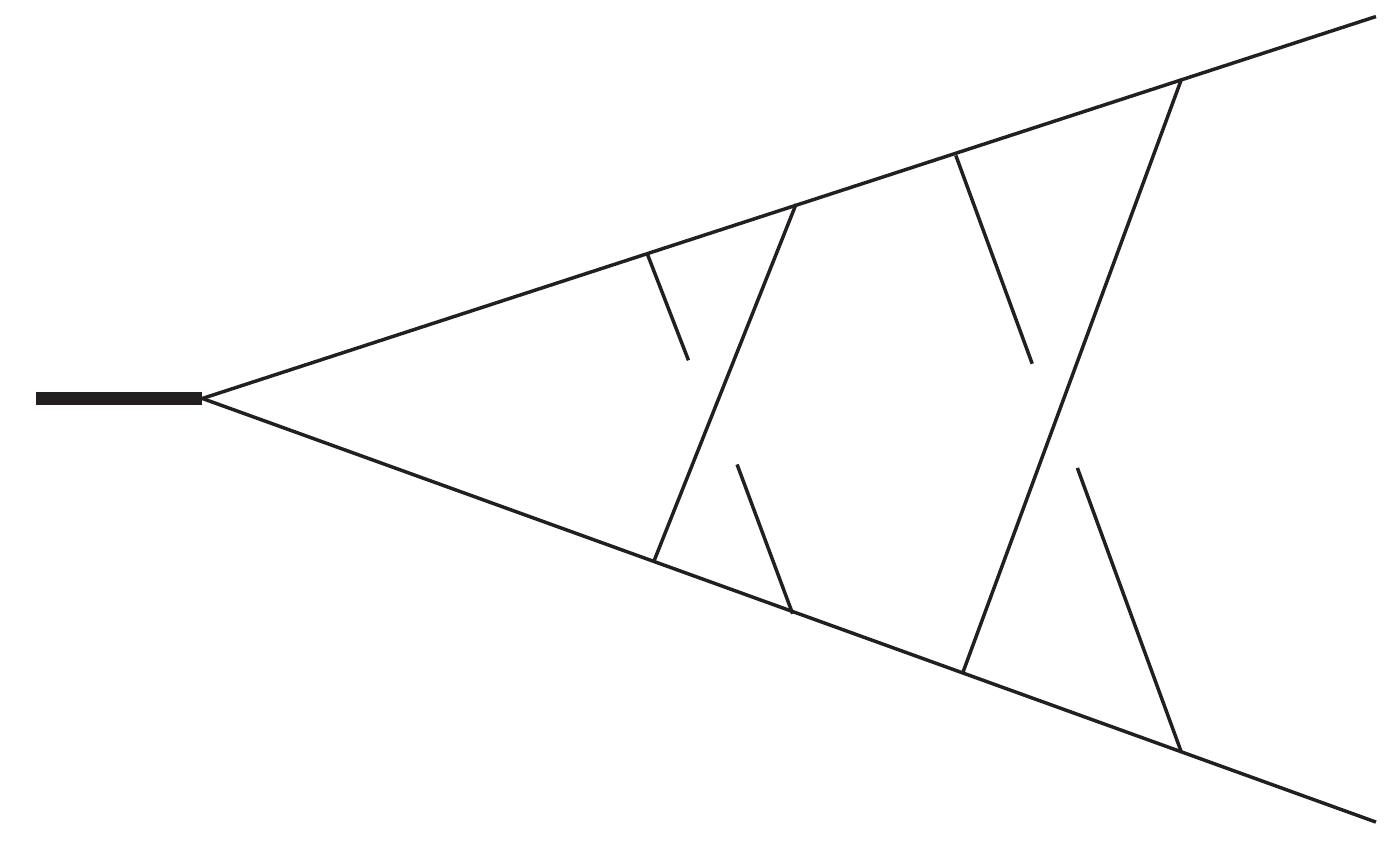}\hspace{-1ex}   
\LabeledGraph{(12)}{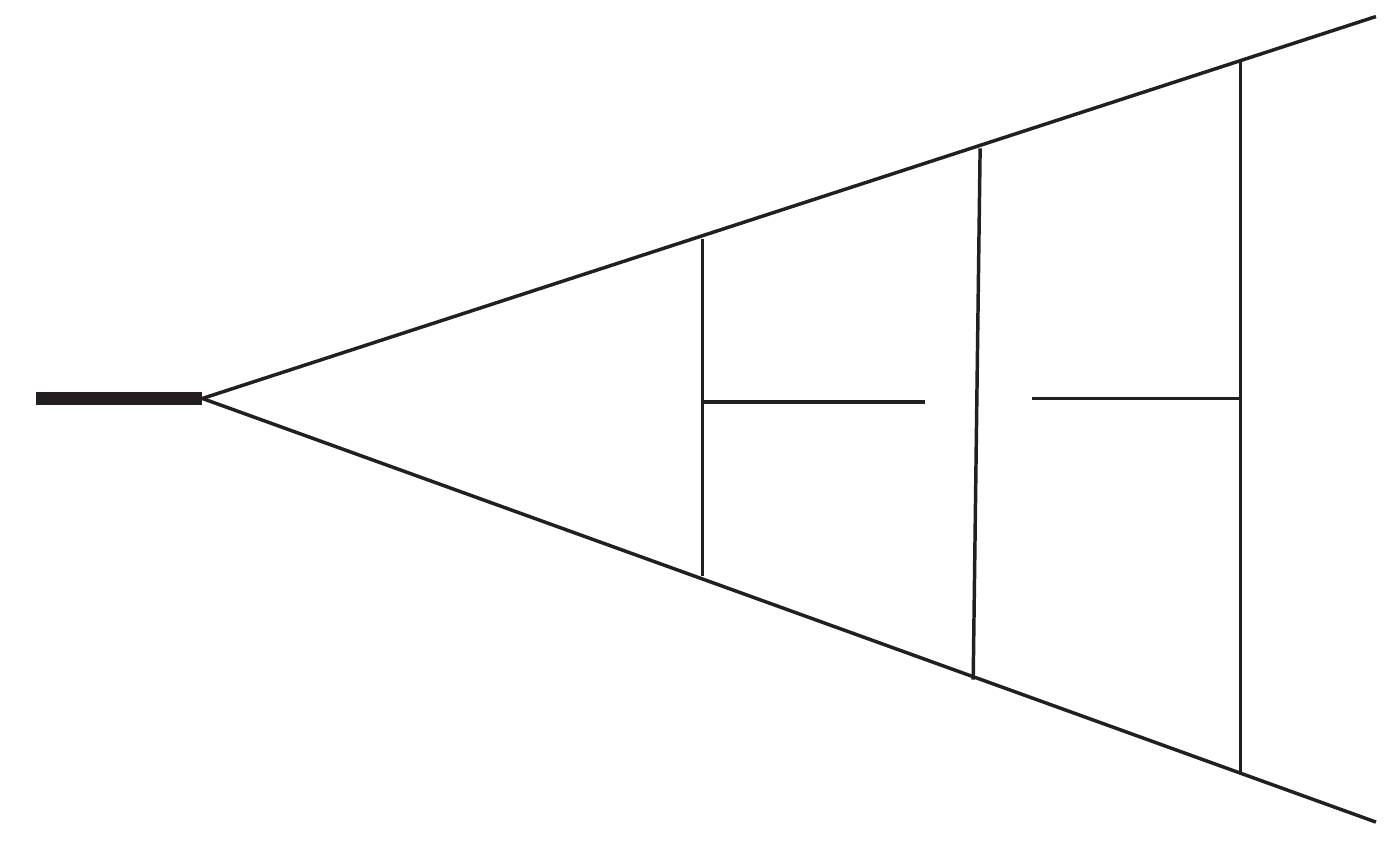}\hspace{-1ex} 
\LabeledGraph{(13)}{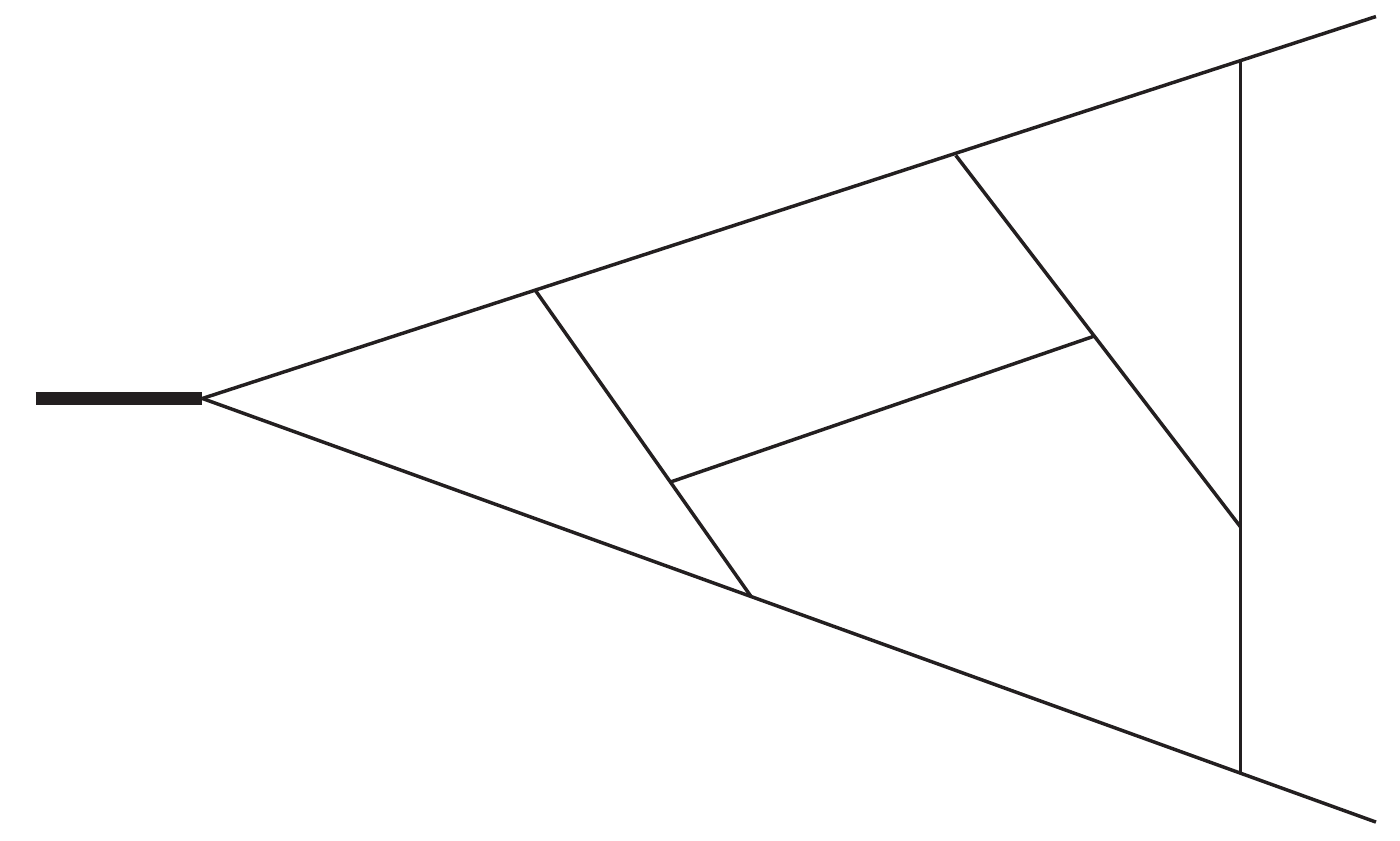}\hspace{-1ex} 
\LabeledGraph{(14)}{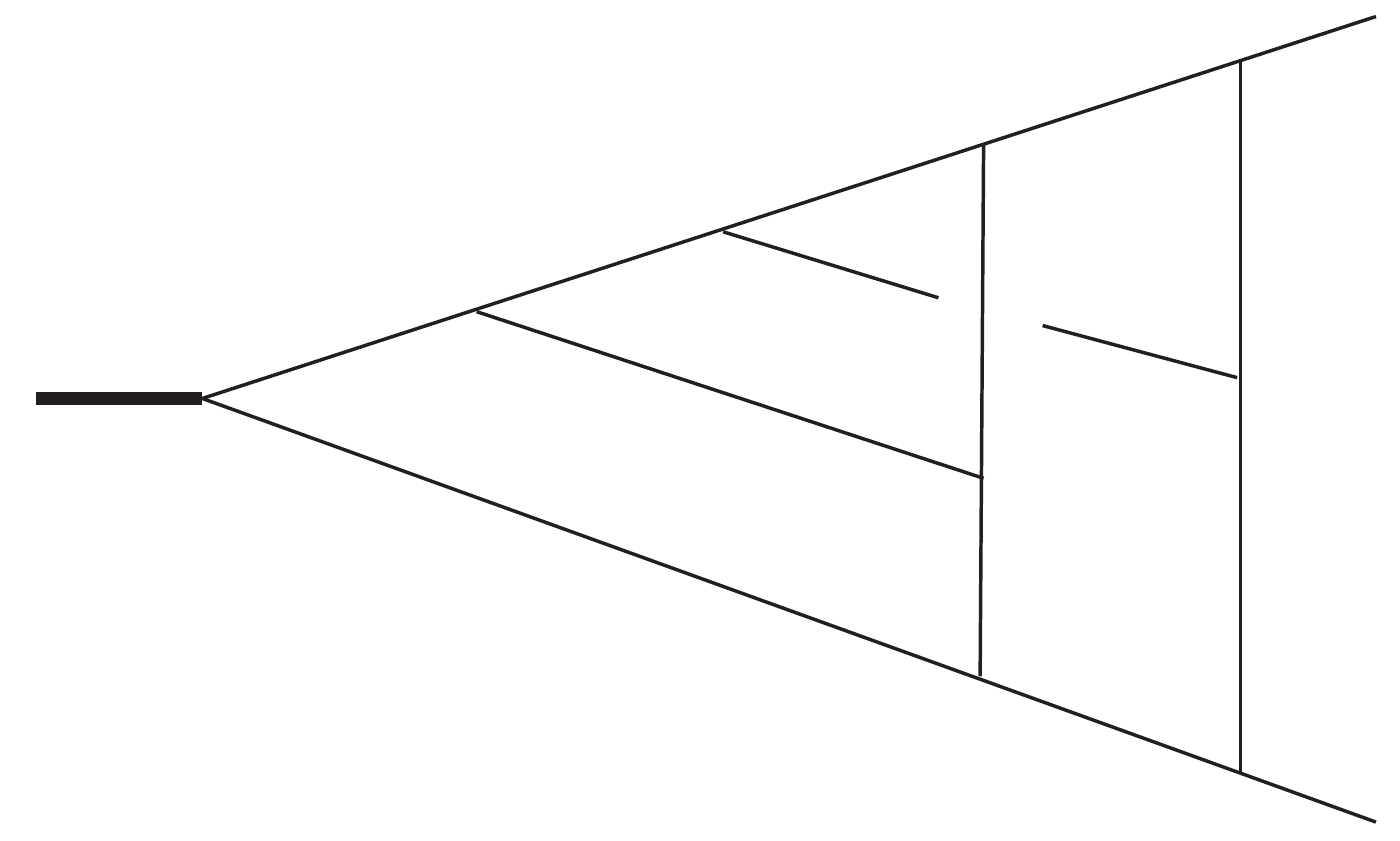}\\[.5ex] 
\LabeledGraph{(17)}{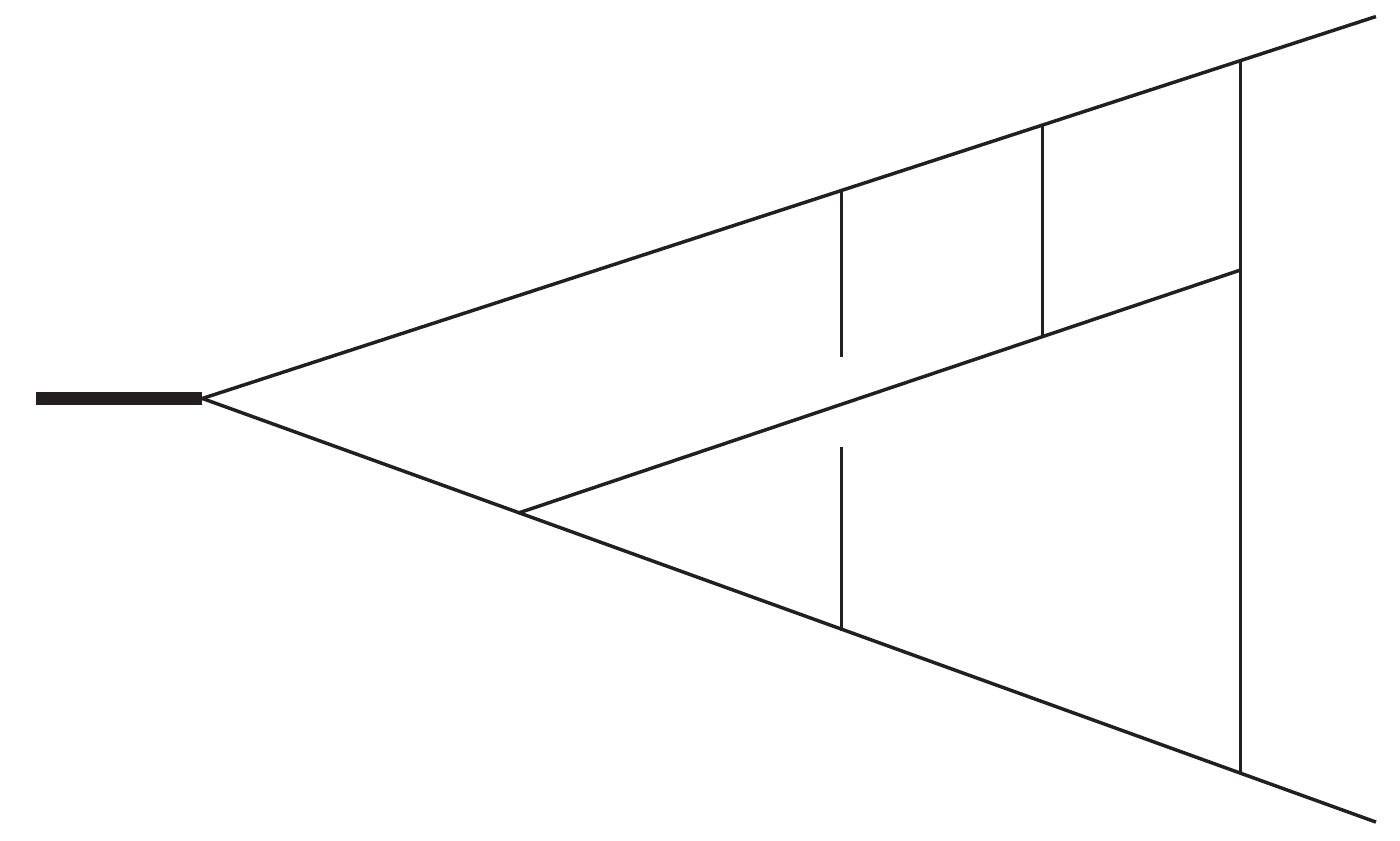}\hspace{-1ex}  
\LabeledGraph{(19)}{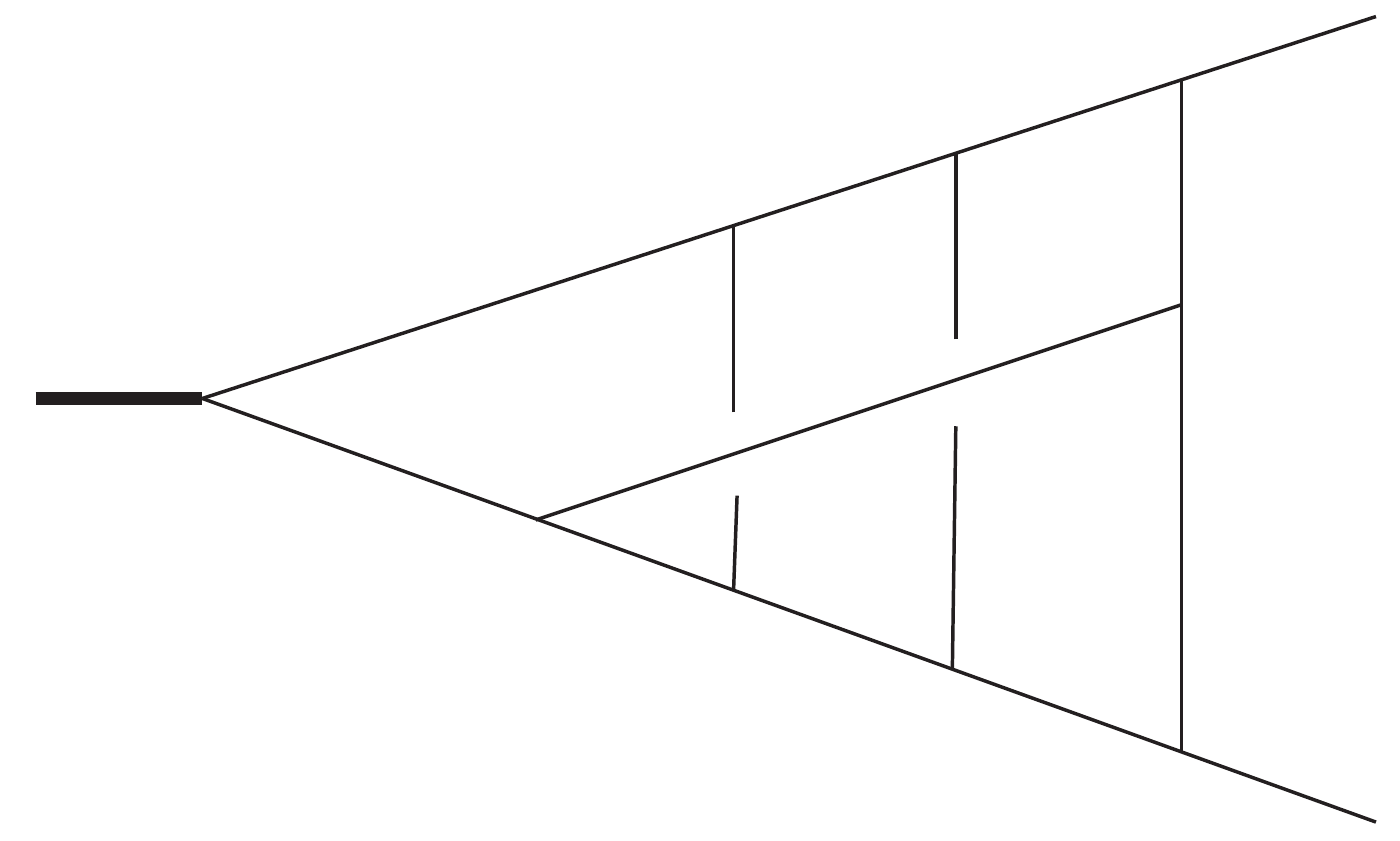}\hspace{-1ex} 
\LabeledGraph{(21)}{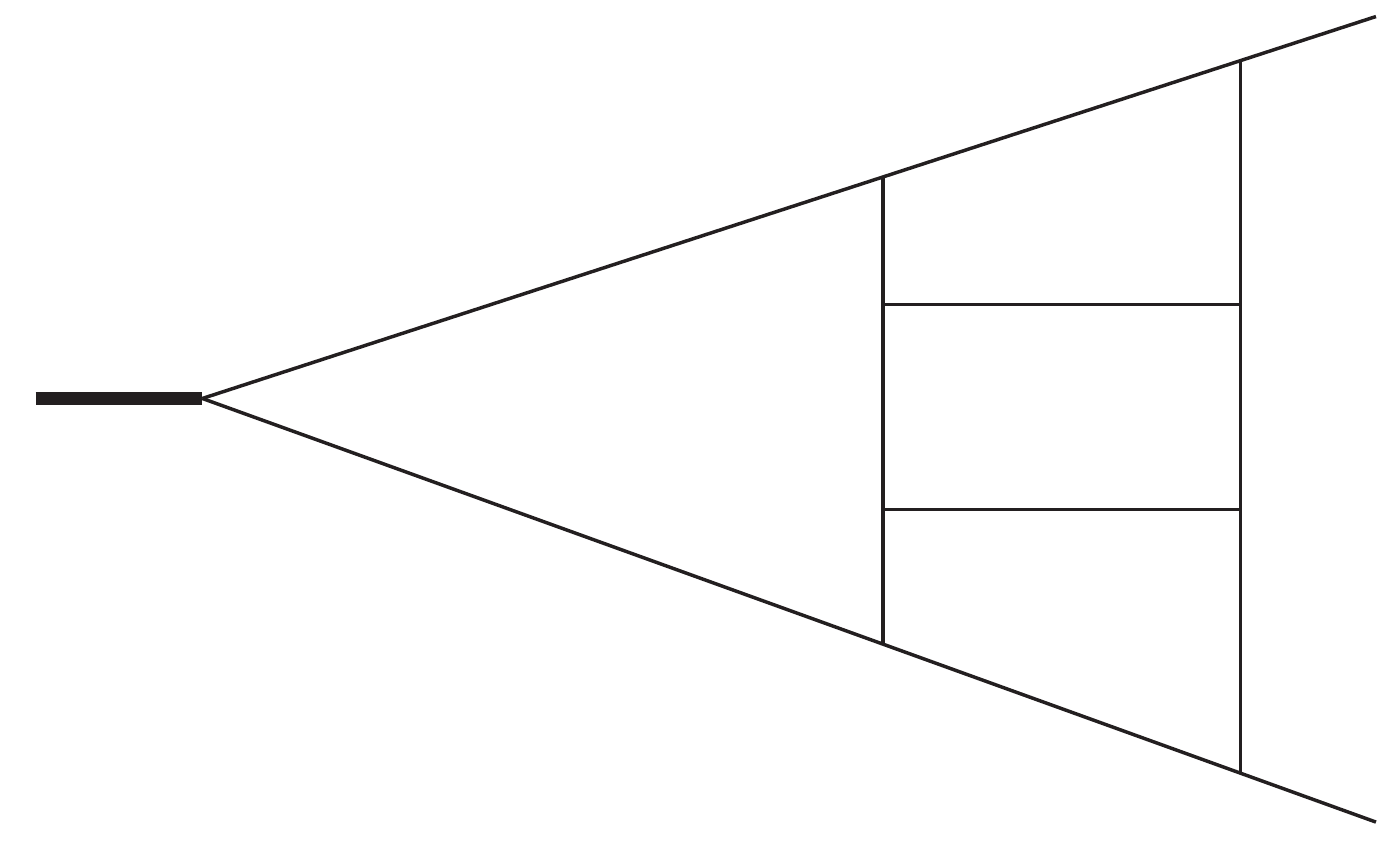}\hspace{-1ex} 
\LabeledGraph{(22)}{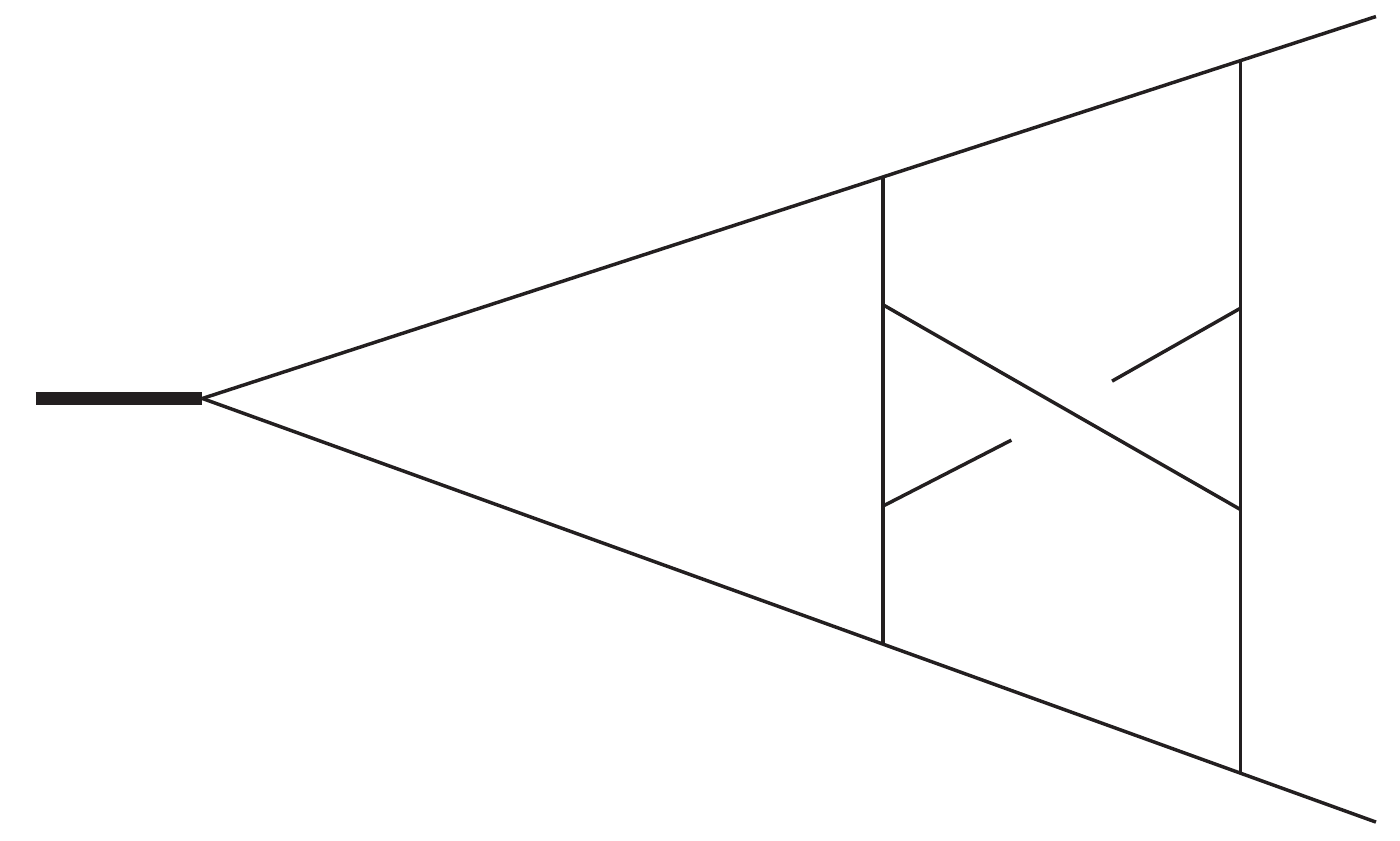}\\[.5ex] 
\LabeledGraph{(23)}{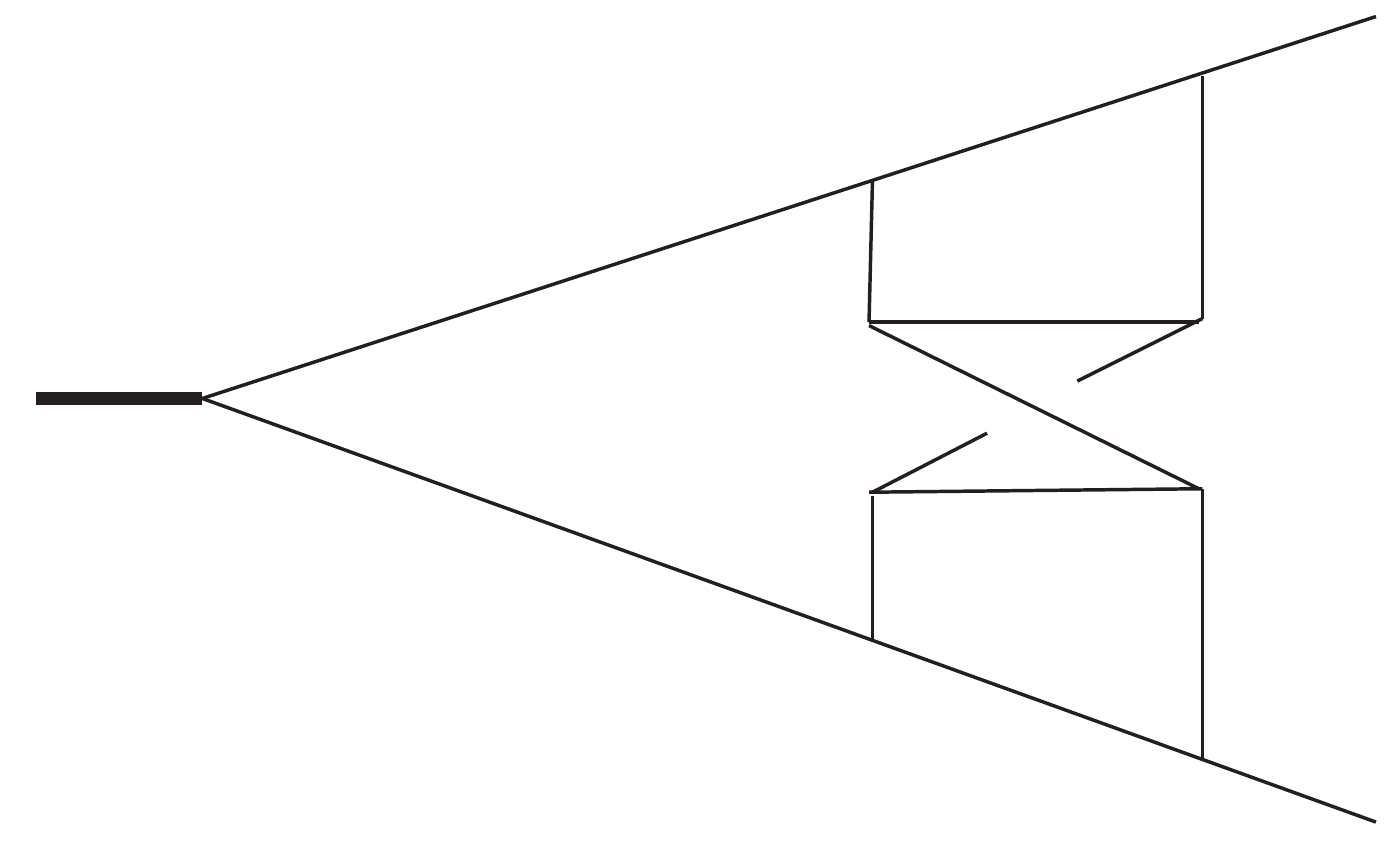}\hspace{-1ex} 
\LabeledGraph{(24)}{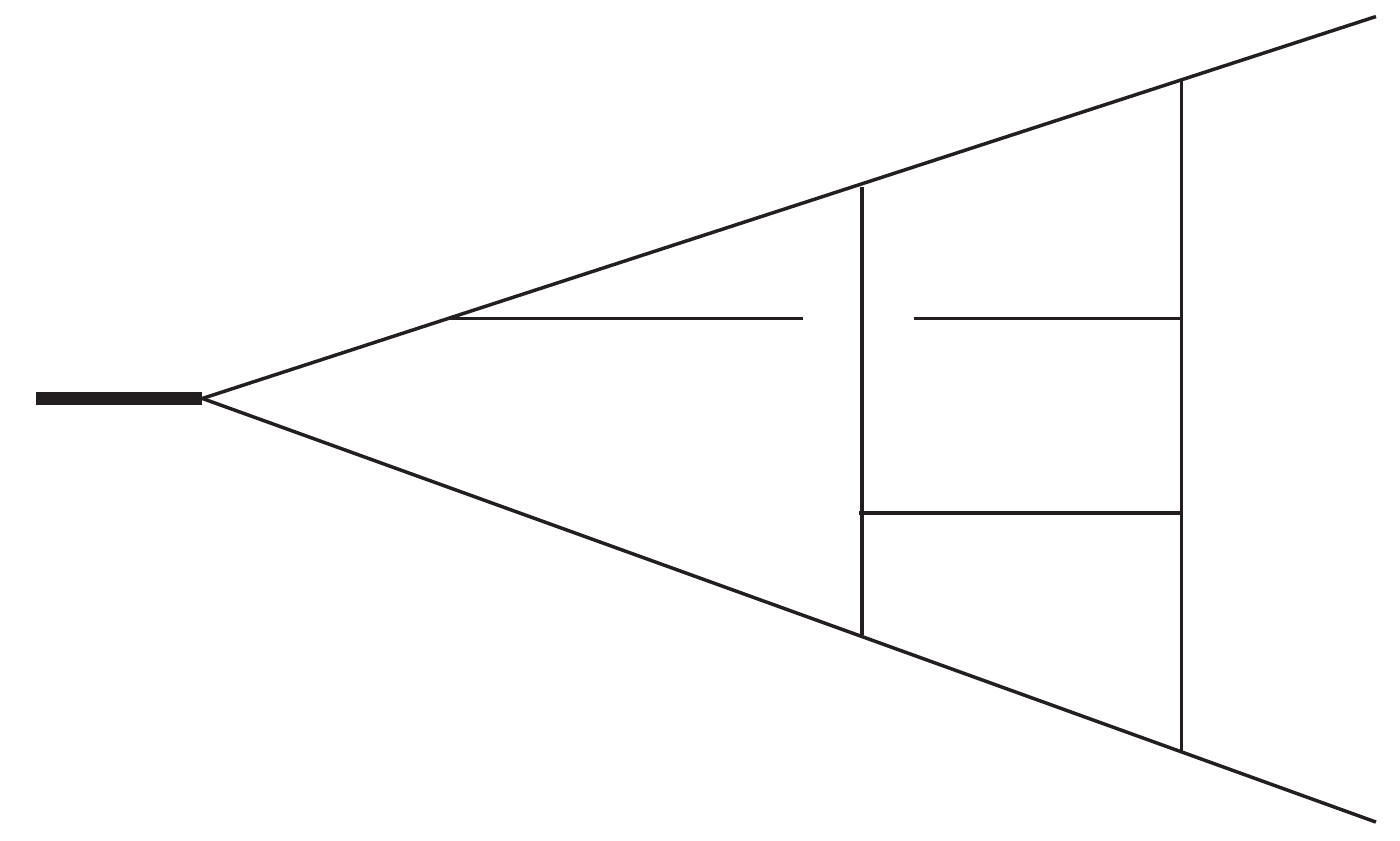}\hspace{-1ex} 
\LabeledGraph{(25)}{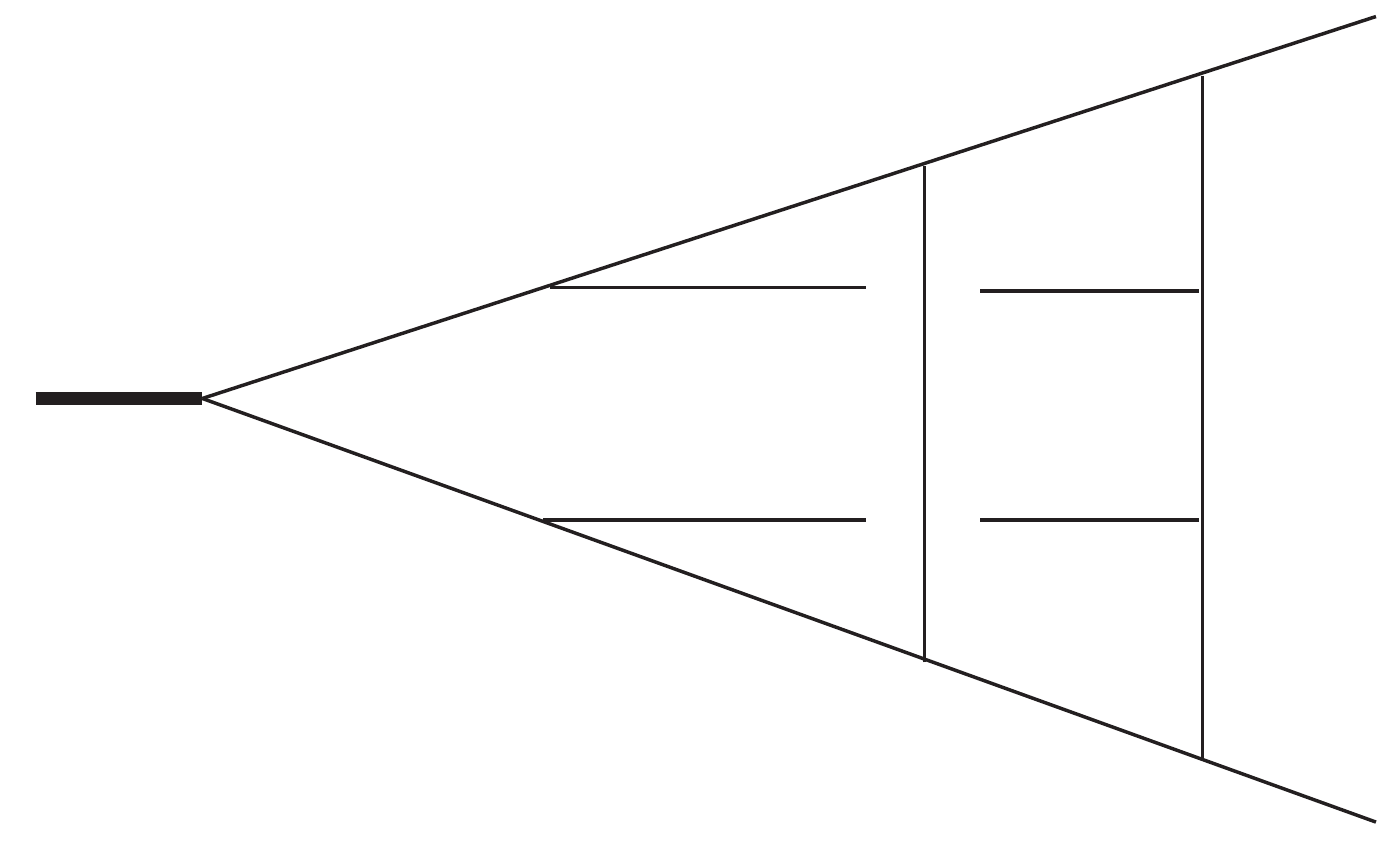}\hspace{-1ex} 
\LabeledGraph{(26)}{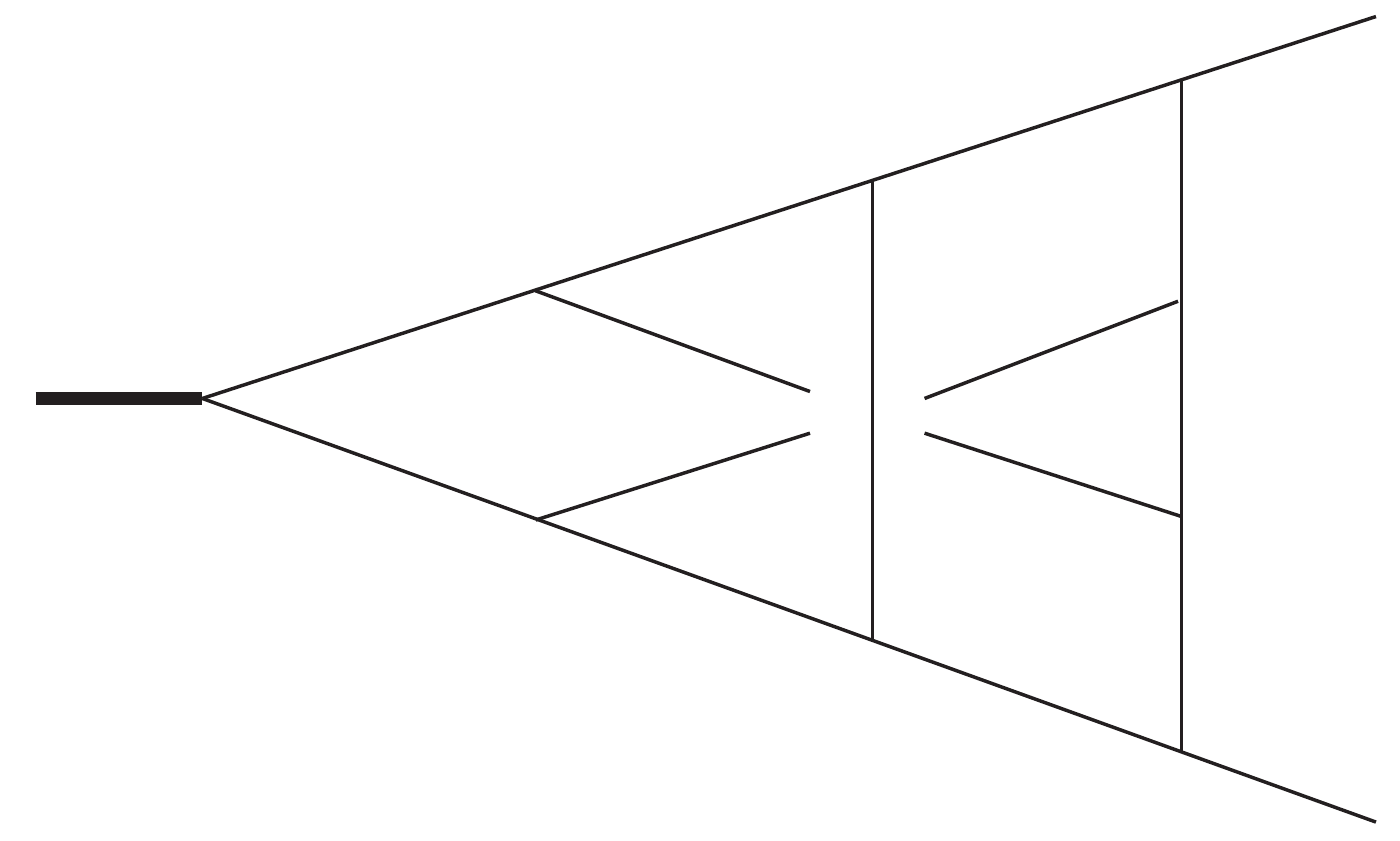}\\[.5ex] 
\LabeledGraph{(27)}{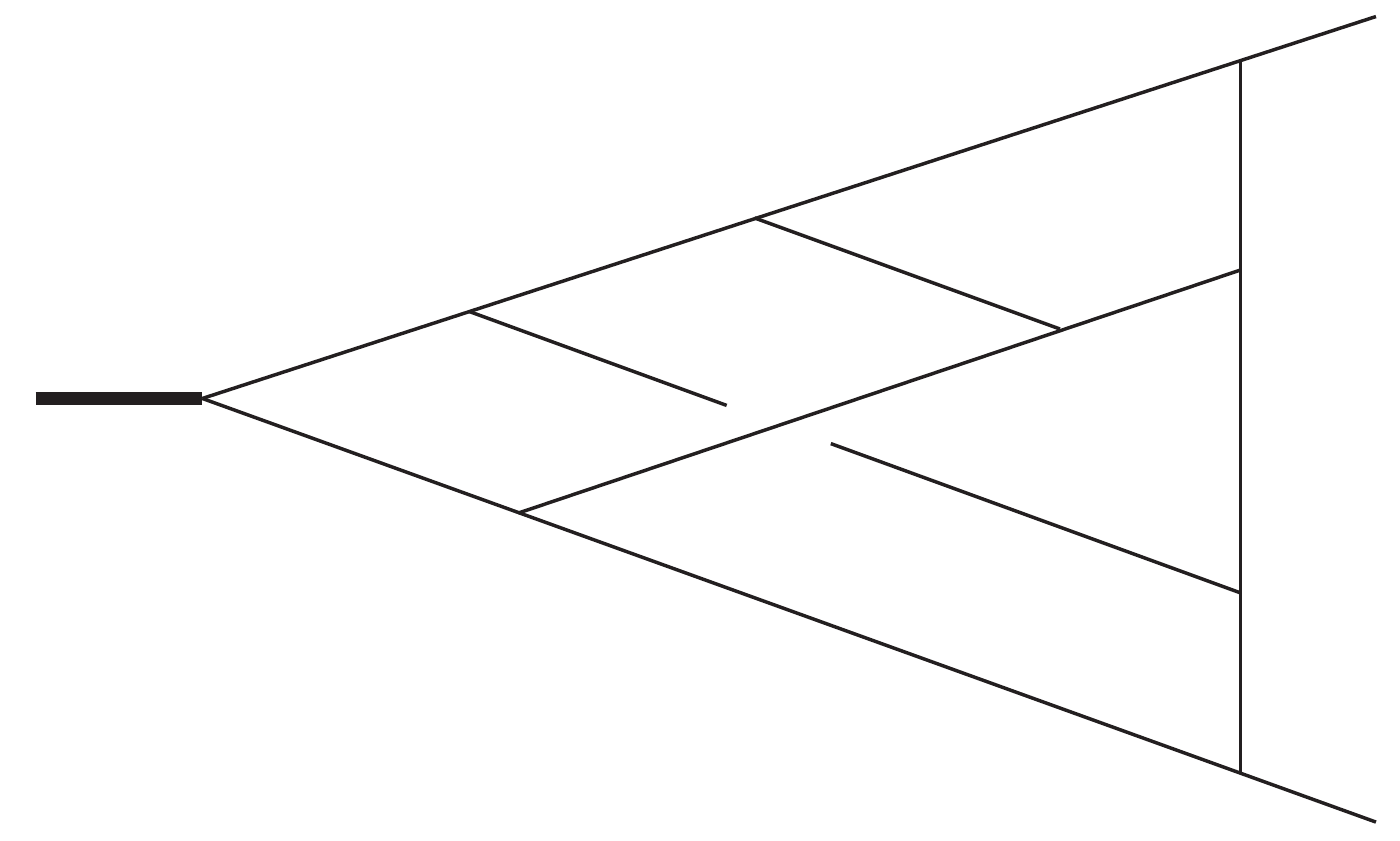}\hspace{-1ex} 
\LabeledGraph{(28)}{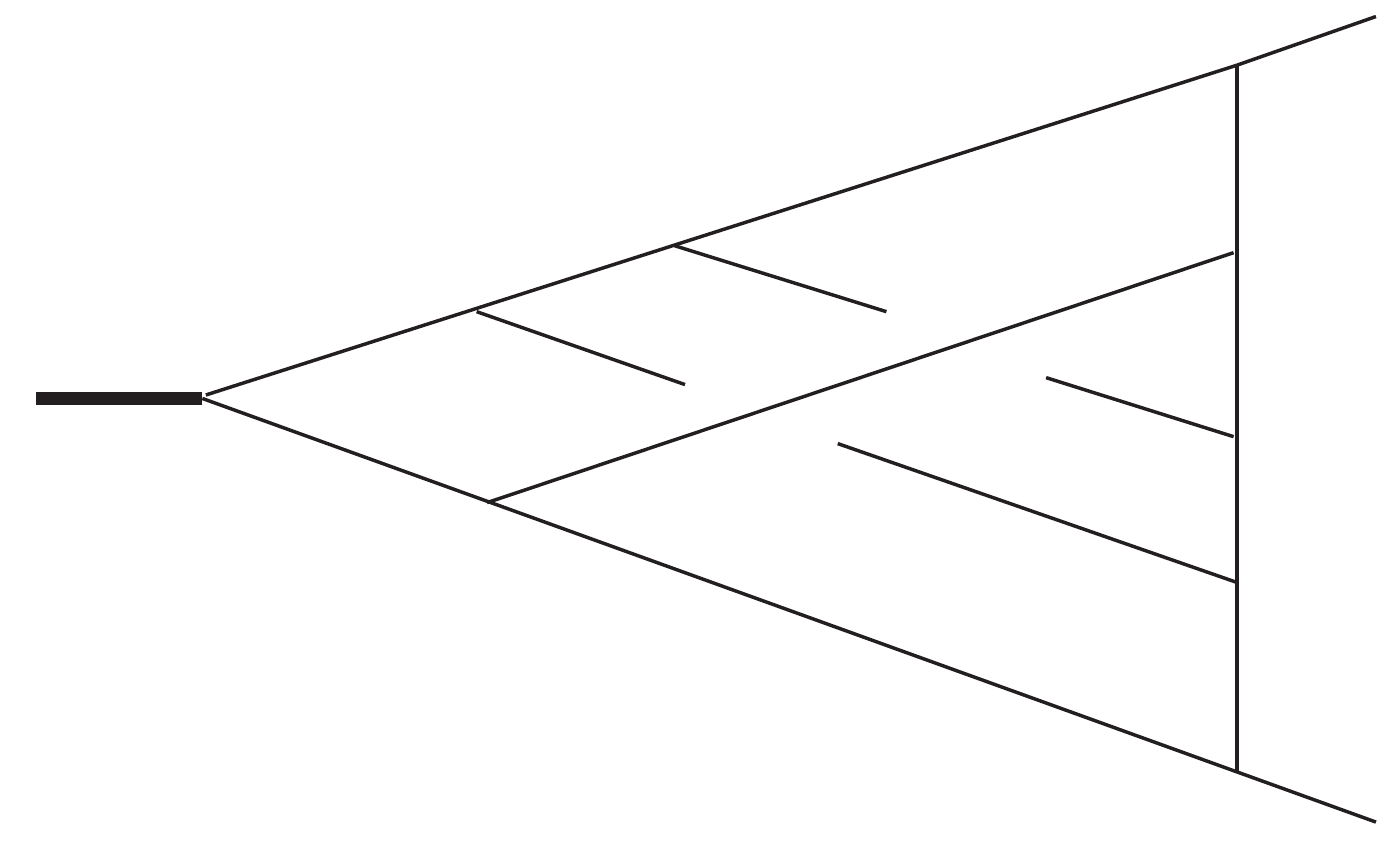}\hspace{-1ex} 
\LabeledGraph{(29)}{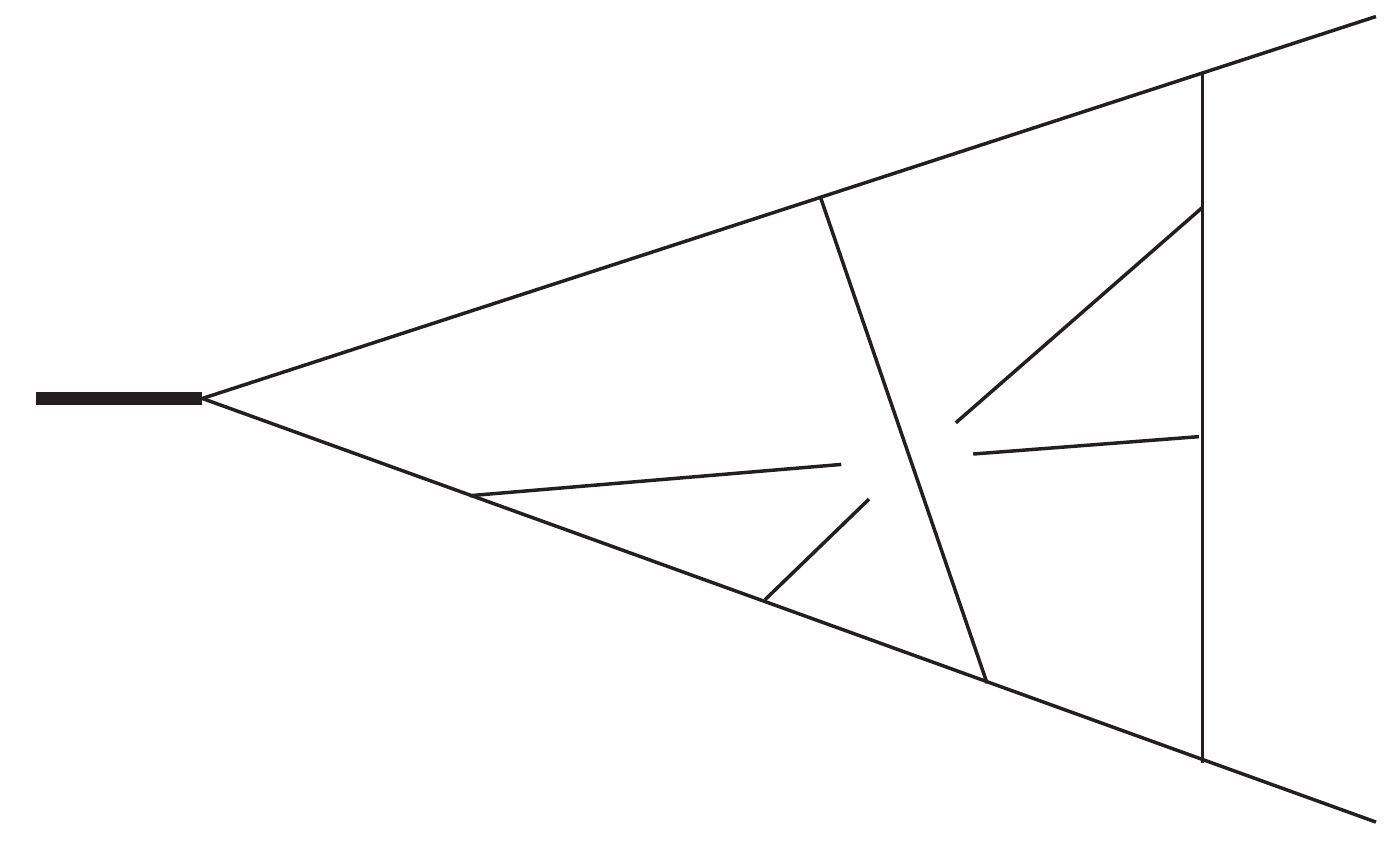}\hspace{-1ex} 
\LabeledGraph{(30)}{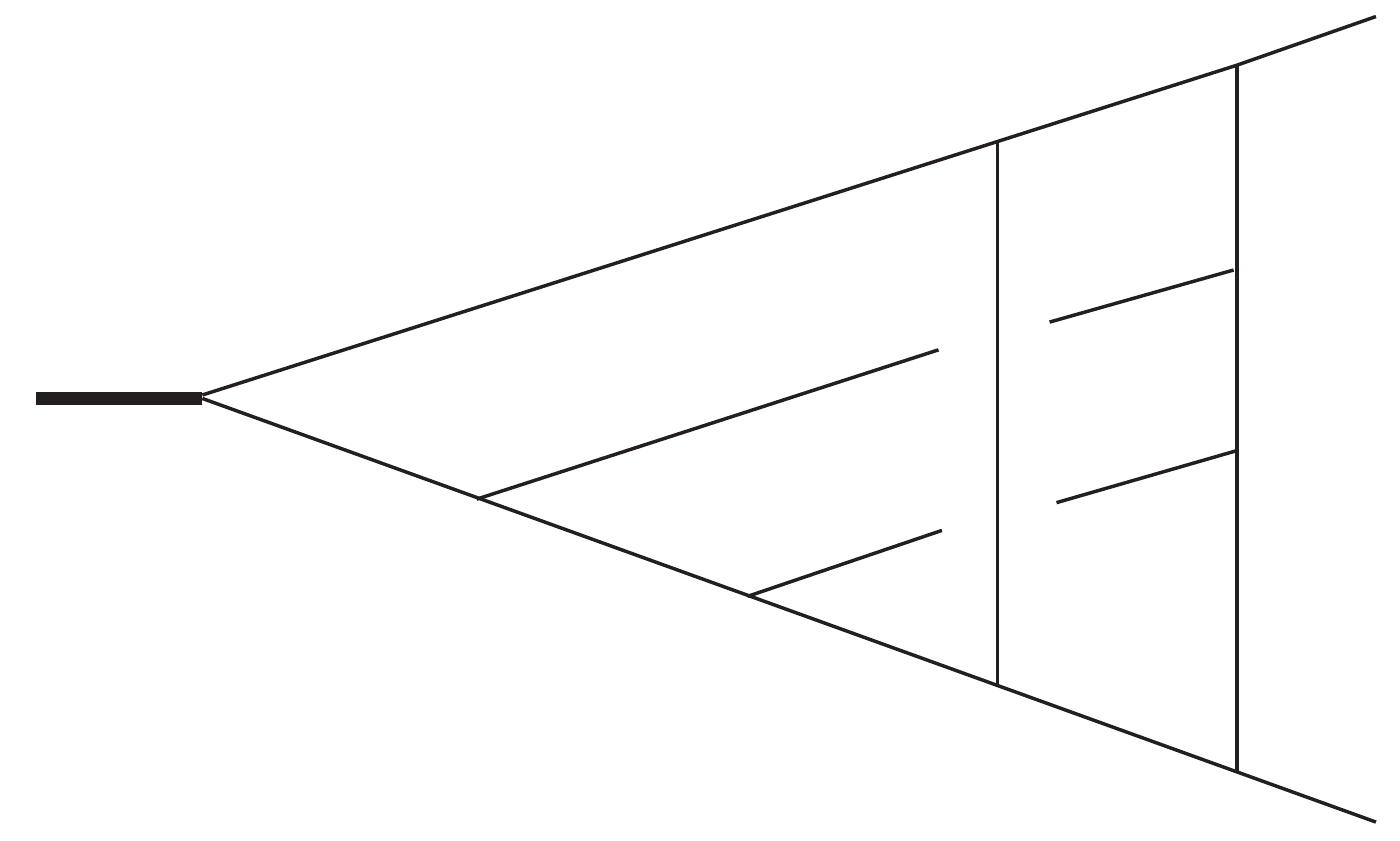} 
\caption{Irreducible trivalent topologies for the $\mathcal{N}=4$ Sudakov form factor at four loops.}
\label{fig:topos}
\end{figure}

The form factor $F$ has been calculated to two loops in \cite{vanNeerven:1985ja} and to three loops in \cite{Gehrmann:2011xn}.
At four loops, a reduced expression for the form factor in terms of dimensionally regularized master integrals has been presented in~\cite{Boels:2017ftb}, which we reproduce here:
\begin{align}
\label{eq:integrand}
 &F^{(4)} = 2\Bigg[8 I_{\mathrm{p}, 1}^{(1)}+2 I_{\mathrm{p}, 2}^{(2)}-2 I_{\mathrm{p}, 3}^{(3)}+2 I_{\mathrm{p}, 4}^{(4)}+\frac{1}{2} I_{\mathrm{p}, 5}^{(5)}+2 I_{\mathrm{p}, 6}^{(6)}+4 I_{\mathrm{p}, 7}^{(7)}+2 I_{\mathrm{p}, 8}^{(9)}-2 I_{\mathrm{p}, 9}^{(10)}+ I_{\mathrm{p}, 10}^{(12)}
 \nonumber \\
 &+ I_{\mathrm{p}, 11}^{(12)}+2 I_{\mathrm{p}, 12}^{(13)}+2 I_{\mathrm{p}, 13}^{(14)}-2 I_{\mathrm{p}, 14}^{(17)}+2 I_{\mathrm{p}, 15}^{(17)}-2 I_{\mathrm{p}, 16}^{(19)}+ I_{\mathrm{p}, 17}^{(19)}+ I_{\mathrm{p}, 18}^{(21)}+\frac{1}{2} I_{\mathrm{p}, 19}^{(25)}+2 I_{\mathrm{p}, 20}^{(30)}+2 I_{\mathrm{p}, 21}^{(13)}
 \nonumber \\
 &+4 I_{\mathrm{p}, 22}^{(14)}-2 I_{\mathrm{p}, 23}^{(14)}- I_{\mathrm{p}, 24}^{(14)}+4 I_{\mathrm{p}, 25}^{(17)}- I_{\mathrm{p}, 26}^{(17)}-2 I_{\mathrm{p}, 27}^{(17)}-2 I_{\mathrm{p}, 28}^{(17)}- I_{\mathrm{p}, 29}^{(19)}
- I_{\mathrm{p}, 30}^{(19)}+ I_{\mathrm{p}, 31}^{(19)}-\frac{1}{2} I_{\mathrm{p}, 32}^{(30)}\Bigg]
 \nonumber \\
 &+\frac{48}{N_c^2}\Bigg[ \frac{1}{2} I_{1}^{(21)}+\frac{1}{2} I_{2}^{(22)}+\frac{1}{2} I_{3}^{(23)}- I_{4}^{(24)}+\frac{1}{4} I_{5}^{(25)}-\frac{1}{4} I_{6}^{(26)}-\frac{1}{4} I_{7}^{(26)}+2 I_{8}^{(27)}+I_{9}^{(28)}
 \nonumber \\
 &+4 I_{10}^{(29)}+ I_{11}^{(30)}+ I_{12}^{(27)}-\frac{1}{2} I_{13}^{(28)}+ I_{14}^{(29)}+ I_{15}^{(29)}+ I_{16}^{(30)}+ I_{17}^{(30)}+ I_{18}^{(30)}+ I_{19}^{(22)}+ I_{20}^{(22)}
 \nonumber \\
 &- I_{21}^{(24)}+\frac{1}{4} I_{22}^{(24)}+\frac{1}{2} I_{23}^{(28)}\Bigg].
\end{align}
Three integrals appear in both the planar-color and non-planar-color parts,
\begin{equation}
I_{1}^{(21)} = I_{\mathrm{p}, 18}^{(21)} , \qquad 
I_{5}^{(25)} = I_{\mathrm{p}, 19}^{(25)} , \qquad
I_{11}^{(30)} = I_{\mathrm{p}, 20}^{(30)}.
\end{equation}
The definition of the integrals $I^{(n_i)}_i$ in Eq.~\eqref{eq:integrand} in terms of propagators for the topology $n_i$ can be found in \cite{Boels:2017ftb}.
Figure~\ref{fig:topos} shows a subset of the contributing topologies: the set of twelve-line topologies with master integrals.
We note that the ``planar-color'' part of the form factor involves both planar and non-planar topologies.
A remarkable feature of Eq.~\eqref{eq:integrand} is that the coefficients of the master integrals are just rational numbers and all dependence on $\epsilon$ is encapsulated in the master integrals, which are conjectured to be of uniform transcendental weight.

\begin{table}
\centering\scalebox{0.71}{ 
\centering\begin{tabular}{|c|c|c|c|c|c|}
\hline & A & B & C & D & E\\ \hline
$D_{1}$ & $k_1^2$ & $(k_1 \!+\! k_2 \!-\! k_3 \!-\! k_4 \!-\! p_1)^2$ \
& $(k_1 \!-\! k_3 \!-\! p_1)^2$ & $(k_1 \!+\! k_2 \!-\! k_3 \!-\! k_4 \
\!-\! p_1)^2$ & $(k_1 \!-\! k_2 \!+\! k_3 \!-\! k_4 \!+\! p_1)^2$\\
$D_{2}$ & $k_2^2$ & $(k_2 \!-\! k_4 \!-\! p_1)^2$ & $(k_3 \!-\! k_4 \
\!+\! p_1)^2$ & $(k_2 \!-\! k_3 \!-\! p_1)^2$ & $(k_1 \!-\! k_2 \!+\! \
k_3 \!+\! p_1)^2$\\
$D_{3}$ & $k_3^2$ & $(k_4 \!+\! p_1)^2$ & $(k_1 \!-\! k_3 \!+\! \
p_2)^2$ & $(k_2 \!-\! p_1)^2$ & $(k_1 \!-\! k_2 \!+\! p_1)^2$\\
$D_{4}$ & $k_4^2$ & $(k_1 \!+\! k_2 \!-\! k_3 \!-\! k_4 \!+\! p_2)^2$ \
& $(k_1 \!-\! k_2 \!+\! p_2)^2$ & $(k_1 \!+\! k_2 \!-\! k_3 \!-\! k_4 \
\!+\! p_2)^2$ & $(k_1 \!+\! p_1)^2$\\
$D_{5}$ & $(k_1 \!-\! p_1)^2$ & $(k_1 \!-\! k_4 \!+\! p_2)^2$ & \
$k_1^2$ & $(k_1 \!-\! k_4 \!+\! p_2)^2$ & $(k_1 \!-\! k_2 \!+\! k_3 \
\!-\! k_4 \!-\! p_2)^2$\\
$D_{6}$ & $(k_1 \!-\! k_2 \!-\! p_1)^2$ & $(k_4 \!-\! p_2)^2$ & \
$k_2^2$ & $(k_1 \!+\! p_2)^2$ & $(k_1 \!-\! k_2 \!+\! k_3 \!-\! \
p_2)^2$\\
$D_{7}$ & $(k_1 \!-\! k_2 \!+\! k_3 \!-\! p_1)^2$ & $k_1^2$ & $k_3^2$ \
& $k_1^2$ & $(k_2 \!-\! k_3 \!+\! p_2)^2$\\
$D_{8}$ & $(k_1 \!-\! k_2 \!+\! k_3 \!-\! k_4 \!-\! p_1)^2$ & $k_2^2$ \
& $k_4^2$ & $k_2^2$ & $k_1^2$\\
$D_{9}$ & $(k_1 \!+\! p_2)^2$ & $k_3^2$ & $(k_2 \!-\! k_3)^2$ & \
$k_3^2$ & $k_2^2$\\
$D_{10}$ & $(k_1 \!-\! k_2 \!+\! p_2)^2$ & $k_4^2$ & $(k_1 \!-\! \
k_2)^2$ & $k_4^2$ & $k_3^2$\\
$D_{11}$ & $(k_1 \!-\! k_2 \!+\! k_3 \!+\! p_2)^2$ & $(k_2 \!-\! \
k_3)^2$ & $(k_3 \!-\! k_4)^2$ & $(k_2 \!-\! k_3)^2$ & $k_4^2$\\
$D_{12}$ & $(k_1 \!-\! k_2 \!+\! k_3 \!-\! k_4 \!+\! p_2)^2$ & $(k_1 \
\!-\! k_3)^2$ & $(k_1 \!-\! k_4)^2$ & $(k_1 \!-\! k_4)^2$ & $(k_2 \!-\! \
k_3)^2$\\
$D_{13}$ & $(k_1 \!-\! k_2)^2$ & $(k_1 \!-\! k_4 \!-\! p_1)^2$ & \
$(k_1 \!-\! k_2 \!-\! p_1)^2$ & $(k_2 \!-\! k_4 \!-\! p_1)^2$ & $(k_3 \
\!-\! p_2)^2$\\
$D_{14}$ & $(k_2 \!-\! k_3)^2$ & $(k_2 \!-\! k_4 \!+\! p_2)^2$ & \
$(k_3 \!-\! k_4 \!-\! p_2)^2$ & $(k_1 \!-\! k_3 \!+\! p_2)^2$ & $(k_1 \
\!-\! k_2)^2$\\
$D_{15}$ & $(k_3 \!-\! k_4)^2$ & $(k_2 \!-\! k_4)^2$ & $(k_1 \!-\! \
p_1)^2$ & $(k_2 \!-\! k_4)^2$ & $(k_1 \!-\! k_3)^2$\\
$D_{16}$ & $(k_1 \!-\! k_2 \!+\! k_3)^2$ & $(k_1 \!-\! k_2)^2$ & \
$(k_1 \!+\! p_2)^2$ & $(k_1 \!-\! k_3)^2$ & $(k_1 \!-\! k_4)^2$\\
$D_{17}$ & $(k_2 \!-\! k_3 \!+\! k_4)^2$ & $(k_1 \!-\! k_4)^2$ & \
$(k_1 \!-\! k_3)^2$ & $(k_3 \!-\! k_4)^2$ & $(k_2 \!-\! k_4)^2$\\
$D_{18}$ & $(k_1 \!-\! k_2 \!+\! k_3 \!-\! k_4)^2$ & $(k_1 \!+\! k_2 \
\!-\! k_3 \!-\! k_4)^2$ & $(k_2 \!-\! k_4)^2$ & $(k_1 \!-\! k_2)^2$ & \
$(k_3 \!-\! k_4)^2$\\
\hline
\end{tabular}
}
\\[2ex]
\centering\scalebox{0.71}{ 
\!\!\begin{tabular}{|c|c|c|c|c|c|}
\hline & F & G & H & I & J\\ \hline
$D_{1}$ & $(k_1 \!+\! k_2 \!-\! k_3 \!-\! k_4 \!-\! p_1)^2$ & $(k_1 \
\!-\! k_2 \!-\! k_3 \!+\! k_4 \!-\! p_1)^2$ & $(k_1 \!-\! p_1)^2$ & \
$(k_1 \!+\! k_3 \!-\! k_4 \!-\! p_1)^2$ & $k_1^2$\\
$D_{2}$ & $(k_1 \!+\! k_2 \!-\! k_4 \!-\! p_1)^2$ & $(k_1 \!-\! k_2 \
\!+\! k_4 \!-\! p_1)^2$ & $(k_1 \!+\! k_2 \!-\! p_1)^2$ & $(k_3 \!-\! \
k_4 \!-\! p_1)^2$ & $k_2^2$\\
$D_{3}$ & $(k_2 \!-\! p_1)^2$ & $(k_1 \!-\! k_2 \!-\! p_1)^2$ & $(k_1 \
\!+\! k_2 \!-\! k_3 \!-\! p_1)^2$ & $(k_4 \!+\! p_1)^2$ & $k_3^2$\\
$D_{4}$ & $(k_1 \!+\! k_2 \!-\! k_3 \!-\! k_4 \!+\! p_2)^2$ & $(k_1 \
\!-\! k_2 \!-\! k_3 \!+\! k_4 \!+\! p_2)^2$ & $(k_1 \!+\! k_2 \!-\! \
k_3 \!-\! k_4 \!-\! p_1)^2$ & $(k_2 \!-\! k_4 \!-\! p_1)^2$ & $k_4^2$\\
$D_{5}$ & $(k_1 \!-\! k_3 \!+\! p_2)^2$ & $(k_2 \!-\! k_4 \!-\! \
p_2)^2$ & $(k_2 \!+\! p_2)^2$ & $(k_1 \!+\! k_3 \!-\! k_4 \!+\! \
p_2)^2$ & $(k_1 \!+\! p_1)^2$\\
$D_{6}$ & $(k_1 \!+\! p_2)^2$ & $k_3^2$ & $(k_1 \!+\! k_2 \!+\! \
p_2)^2$ & $(k_1 \!-\! k_4 \!+\! p_2)^2$ & $(k_1 \!-\! k_3 \!+\! \
p_1)^2$\\
$D_{7}$ & $k_1^2$ & $k_4^2$ & $(k_1 \!+\! k_2 \!-\! k_3 \!+\! p_2)^2$ \
& $(k_4 \!-\! p_2)^2$ & $(k_1 \!+\! k_2 \!-\! k_3 \!+\! p_1)^2$\\
$D_{8}$ & $k_2^2$ & $(k_1 \!-\! k_2)^2$ & $(k_1 \!+\! k_2 \!-\! k_3 \
\!-\! k_4 \!+\! p_2)^2$ & $k_1^2$ & $(k_1 \!+\! k_2 \!-\! k_3 \!-\! \
k_4 \!+\! p_1)^2$\\
$D_{9}$ & $k_3^2$ & $(k_2 \!-\! k_3)^2$ & $k_1^2$ & $k_2^2$ & $(k_3 \
\!+\! p_2)^2$\\
$D_{10}$ & $(k_1 \!-\! k_2)^2$ & $(k_2 \!-\! k_4)^2$ & $k_2^2$ & \
$k_3^2$ & $(k_1 \!-\! k_3 \!-\! p_2)^2$\\
$D_{11}$ & $(k_2 \!-\! k_4)^2$ & $(k_3 \!-\! k_4)^2$ & $k_3^2$ & \
$k_4^2$ & $(k_1 \!-\! k_3 \!-\! k_4 \!-\! p_2)^2$\\
$D_{12}$ & $(k_1 \!-\! k_4)^2$ & $(k_1 \!-\! k_3)^2$ & $k_4^2$ & \
$(k_2 \!-\! k_4)^2$ & $(k_1 \!+\! k_2 \!-\! k_3 \!-\! k_4 \!-\! \
p_2)^2$\\
$D_{13}$ & $(k_2 \!-\! k_4 \!-\! p_1)^2$ & $(k_2 \!+\! p_1)^2$ & \
$(k_1 \!-\! k_2)^2$ & $(k_2 \!-\! k_4 \!+\! p_2)^2$ & $(k_1 \!-\! \
k_2)^2$\\
$D_{14}$ & $(k_1 \!+\! k_2 \!-\! k_3 \!+\! p_2)^2$ & $(k_1 \!-\! k_2 \
\!+\! k_4 \!+\! p_2)^2$ & $(k_1 \!-\! k_3)^2$ & $(k_1 \!-\! k_2)^2$ & \
$(k_1 \!-\! k_3)^2$\\
$D_{15}$ & $k_4^2$ & $(k_2 \!-\! p_2)^2$ & $(k_1 \!-\! k_4)^2$ & \
$(k_1 \!-\! k_3)^2$ & $(k_1 \!-\! k_4)^2$\\
$D_{16}$ & $(k_3 \!-\! k_4)^2$ & $k_1^2$ & $(k_2 \!-\! k_3)^2$ & \
$(k_1 \!-\! k_4)^2$ & $(k_2 \!-\! k_3)^2$\\
$D_{17}$ & $(k_1 \!-\! k_3)^2$ & $k_2^2$ & $(k_2 \!-\! k_4)^2$ & \
$(k_2 \!-\! k_3)^2$ & $(k_2 \!-\! k_4)^2$\\
$D_{18}$ & $(k_2 \!-\! k_3)^2$ & $(k_1 \!-\! k_4)^2$ & $(k_3 \!-\! \
k_4)^2$ & $(k_3 \!-\! k_4)^2$ & $(k_3 \!-\! k_4)^2$\\
\hline
\end{tabular}
}
    \caption{A complete set of integral families for massless three-point functions with one off-shell leg at four loops.}
    \label{tab:families}
\end{table}

The definitions of the master integrals in Ref.~\cite{Boels:2017ftb} can be mapped to just the 10 integral families (complete sets of propagators) shown in Table~\ref{tab:families}.
The integrals are then linear combinations of four-loop Feynman integrals
\begin{equation}
\label{eq:intnorm}
    I_{f}(\nu_1,\ldots,\nu_{18}) = (-q^2 e^{\gamma_\text{E}})^{4\ep} 
    \int \left( \prod_{L=1}^4 \frac{\ud^d k_L}{i \pi^{d/2}}\right) 
    \frac{1}{D_1^{\nu_1}\cdots D_{18}^{\nu_{18}}}
\end{equation}
where $f$=A,\ldots,J labels the family and $q^2=-1$.
Depending on the topology, up to 12 indices $\nu_i$ in Eq.~\eqref{eq:intnorm} are positive and correspond to actual denominators of the integrand;  some of the remaining indices may be negative to denote irreducible numerators.
We note that one family covers in general more than one trivalent graph, family A for example covers all planar graphs.
The propagator denominators $D_i$ follow Minkowskian conventions and depend implicitly on the integral family $f$.
We provide expressions for the master integrals $I^{(n_i)}_i$ in terms of integrals in these families in the ancillary files of this paper.
We note that integration-by-parts reductions allow to remove any reference to family I, which was used to map $I_{p,2}^{(2)}$.
Analytical results for the master integrals have been given through to weight 6 in \cite{Huber:2019fxe}.
Here, we present their analytical calculation through to weight 8 as required for the finite part of the Sudakov form factor.

\section{Master integrals to weight eight}
\label{sec:ffmasters}

We employ two different methods to evaluate the master integrals:
the direct integration of finite integrals and the method of differential equations with an auxiliary scale.

In principle, all topologies but two have been shown to be linearly reducible \cite{Brown:2008um,Brown:2009ta} and are thus accessible to direct integrations based on the Feynman parametric representation.
Moreover, the only two topologies which have not yet been proven to be linearly reducible after a change of variables in the Feynman parametric representation were dealt with in Ref.~\cite{Lee:2021uqq} using the method of differential equations.
In order to perform parametric integrations, we select a basis of finite integrals \cite{Panzer:2014gra,vonManteuffel:2014qoa,vonManteuffel:2015gxa,Schabinger:2018dyi,Agarwal:2020dye} with \texttt{Reduze\;2}~\cite{vonManteuffel:2012np}.
Here, the finite integrals are typically defined in $6-2\ep$ dimensions and involve higher powers of the propagators (``dots'').
The basis change is computed with the private code \texttt{Finred} based on~\cite{Tkachov:1981wb,Chetyrkin:1981qh,Laporta:2001dd,vonManteuffel:2014ixa,Lee:2013mka,Bitoun:2017nre}.
For some topologies, it is necessary to perform variable changes in the Feynman parametric representation to find a linearly reducible integration order.
We then employ the program \texttt{HyperInt}~\cite{Panzer:2014caa} to expand the Feynman integral around $\ep=0$ and integrate the expansion coefficients.
In this way, we solved a subset of the master integrals through to weight 8 with \texttt{HyperInt}.
Depending on the integral, however, we found that the computing resources required to compute the relevant $\ep$ orders can be prohibitive, such that we resorted to the method of differential equations in many cases.

The method of differential equations~\cite{Kotikov:1990kg,Bern:1993kr,Gehrmann:1999as} is a powerful technique to solve Feynman integrals with non-trivial dependence on the kinematics, see {\it e.g.} Ref.~\cite[Section E.8]{Blondel:2018mad} for a recent review.
While our integrals have only a trivial dependence on the kinematics, the method becomes applicable by considering vertex integrals with two off-shell and one massless leg~\cite{Henn:2013nsa} instead.
The differential equations in the auxiliary parameter
\begin{equation}
x=\frac{p_2^2}{(p_1+p_2)^2}
\end{equation}
then connect the sought after vertex integrals with two massless legs ($x=0$) with propagator type integrals ($x=1$) known from Refs.~\cite{Baikov:2010hf,Lee:2011jt}, see Ref.~\cite{Lee:2019zop} for more details.
We employ \texttt{Fire\;6}~\cite{Smirnov:2019qkx} and \texttt{LiteRed}~\cite{Lee:2013mka,Lee:2012cn} to find the differential equations in $x$ for some initial choice of basis.
Subsequently, we apply the method of Ref.~\cite{Lee:2014ioa,Lee2017c} as implemented in \texttt{Libra}~\cite{Lee:2020zfb} to bring the system in $\ep$ form~\cite{Henn:2013pwa}.
At this point, we are forced to introduce algebraic extensions $x_1=\sqrt{x}$, $x_2=\sqrt{x-1/4}$, and $x_3=\sqrt{1/x-1/4}$ in order to secure an $\ep$-form of the differential system. The complete alphabet sufficient for all families consists of the letters
\begin{equation}
  \frac{1}{x},\ \frac{1}{x+1},\ \frac{1}{x-1},\ \frac{1}{x-4},\ \frac{1}{x-1/4},\ \frac{1}{(1-x)x_1},\ \frac{1}{x x_2},\ \frac{1}{x x_3}
\end{equation}
appearing in the derivatives with respect to $x$.
In particular, the letters involving $x_1,\ x_2,\ x_3$ are required for topology $(26)$ in Fig.\ \ref{fig:topos}, while the topologies $(12)$ and $(25)$ contain those involving $x_1,\ x_3$. It turns out that each iterated integral in the results for master integrals contains at most one of $x_1,\, x_2,\,x_3$, so it is always possible to rationalize the weights by passing to the corresponding letter. 

\begin{figure}[t]
    \centering\includegraphics[width=0.5\textwidth]{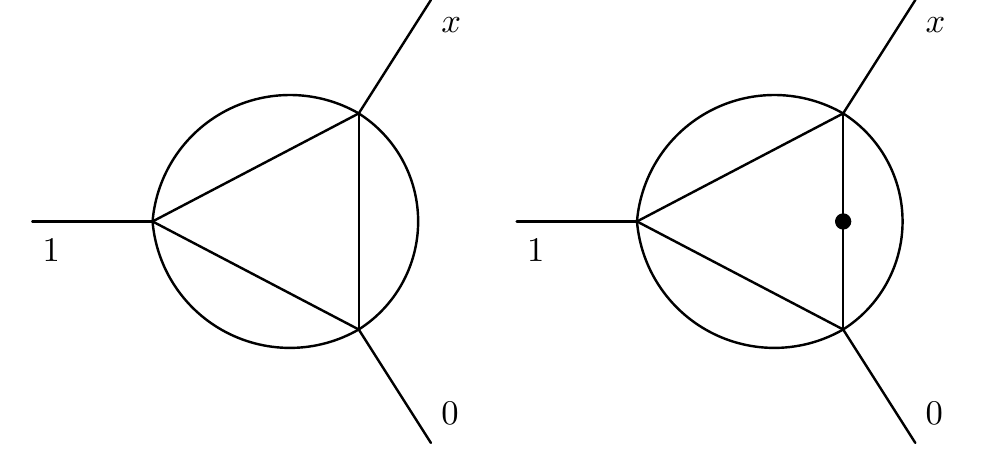}
    \caption{\label{fig:UTexample}Master integrals $j_1$ and $j_2$ of Eq.~(\ref{eq::j1j2}).
    For $x=0$ or $x=1$ the number of master integrals in this sector reduces to one.}
\end{figure}

Note that the differential equations approach allows one to construct uniform transcendentality (UT) bases of one-scale integrals. Indeed, the column of asymptotic coefficients $c_0$ at $x=0$ is expressed via the column of coefficients $c_1$ at $x=1$ as (see Eq.~(28) of Ref.~\cite{Lee:2019zop})
\begin{equation}\label{eq:c0viac1}
    c_0=L_0^{-1}U_{01}L_1c_1 \,.
\end{equation}
Here the associator $U_{01}$ is UT by construction. The column of boundary constants $C_1=L_1c_1$ can also be made UT by a judicious choice of an overall normalization of $L_1$. It suffices to pull from $L_1$ an overall, rational in $\ep$, factor which can be determined by examining the simplest non-zero entry of column $c_1$ (this simplest entry is always known exactly in terms of a product of $\Gamma$ functions). So, the column of boundary constants at $x=0$, \textit{i.e.} $C_0=U_{01}C_1=$ is also UT. On the other hand, from Eq.~\eqref{eq:c0viac1} we have $C_0=L_0c_0$. However, there is one obstacle here. The column $c_0$ contains not only naive limits (obtained by setting $x=0$ under the integral 
sign), which correspond to one-scale integrals, but also the asymptotic coefficients in front of non-integer powers of $x$. Thus, in general, each entry of $C_0$ is expressed not  only via one-scale integrals, but also via some asymptotic coefficients in front of non-integer powers of $x$. This can be fixed by quasi-diagonalizing (reducing to Jordan normal form) the residue,  $A_0$, at $x=0$ of the matrix on the right-hand side of the differential system in $\ep$-form. Since the fractional powers of $x$ in the asymptotics are in one-to-one correspondence with eigenspaces of $A_0$, the Jordan normal form of $A_0$ necessarily has a block-diagonal structure with blocks corresponding to different fractional powers of $x$. The matrix $L_0$ also acquires the same block-diagonal structure. Then, those entries of $C_0$ which correspond to a block with integer powers of $x$ are expressed solely via one-scale integrals. Since the matrix $L_0$ is invertible by construction, it is easy to establish, that the number of such entries is sufficient to furnish a basis. 

Let us demonstrate this approach on the example of the two integrals presented in Fig.~\ref{fig:UTexample}, where we use a dot to indicate a squared propagator.
The differential system for those two integrals has the form
\begin{equation}
\frac{\ud}{\ud x}\begin{pmatrix} j_1 \\j_2 \end{pmatrix}=
    \begin{pmatrix}
        -\frac{2 (2 \epsilon -1)}{x} & \frac{\epsilon  (3 \epsilon -1)}{x (2 \epsilon -1) (5 \epsilon -3)} \\
        \frac{2 (2 \epsilon -1)^2 (5 \epsilon -3)}{(x-1) x (3 \epsilon -1)} & -\frac{(x+1) \epsilon }{(x-1) x}
    \end{pmatrix}
\begin{pmatrix} j_1 \\j_2 \end{pmatrix}\,.
\label{eq::j1j2}
\end{equation}
Note that this sector has no non-zero subsectors.
We construct the transformation $j=TJ$ with $j=(j_1,j_2)$, $J=(J_1,J_2)$, and 
\begin{equation}
    T=f(\epsilon)
    \begin{pmatrix}
            \frac{(1-3 \epsilon)  (1+x^2)+ 2 \epsilon  x}{2 (3-5 \epsilon ) (1-2 \epsilon )^2} & \frac{(1-3 \epsilon )(1-x^2) }{2 (3-5 \epsilon ) (1-2 \epsilon )^2} \\
             \frac{1-3 \epsilon +x \epsilon}{(1-3\epsilon)\epsilon} &\frac{ 1 -3 \epsilon -x \epsilon}{(1-3\epsilon)\epsilon}
    \end{pmatrix}\,
\end{equation}
which reduces the system to an $\ep$-form. The factor $f(\epsilon)$ will be fixed later to secure uniform transcendentality of $C_0$ and $C_1$. We have 
\begin{equation}
    \frac{\ud}{\ud x}\begin{pmatrix} J_1 \\J_2 \end{pmatrix}=
   \epsilon S(x)\begin{pmatrix} J_1 \\J_2 \end{pmatrix},
\end{equation}
where 
\begin{equation}
    S(x)= \frac{S_0}{x} +  \frac{S_1}{x-1},\qquad S_0=\begin{pmatrix} -2 & 1 \\ 2 & -1 \end{pmatrix}
    ,\qquad S_1=\begin{pmatrix} 0 & 0 \\ 0 & -2 \end{pmatrix}\,.\label{eq:S0S1}
\end{equation}
Using \texttt{Libra}, we find the following relation between asymptotic coefficients at $x=0$ and $x=1$:
\begin{equation}
    \begin{pmatrix}
        [j_1]_{x^0} \\ [j_1]_{x^{2-3\epsilon}}
    \end{pmatrix}
=L_0^{-1} U_{01}L_1  
\begin{pmatrix}
    [j_1]_{(1-x)^0} \\ [j_2]_{(1-x)^{-2\epsilon}}
\end{pmatrix}\,,
\end{equation}
where $[j_k]_{y^\alpha}$ denotes the coefficient in front of $y^\alpha$ in $y\to 0$ asymptotics of $j_k$. In principle, also different choices are possible (e.g. $[j_2]_{x^{2-3\epsilon}}$ instead of $[j_1]_{x^{2-3\epsilon}}$) with appropriate modifications of $L_0$ and $L_1$. For our choice the matrices $L_0$ and $L_1$ have the form
\begin{align}
    L_0&=
   f(\epsilon)^{-1}
     \begin{pmatrix}
             \frac{2 (3-5 \epsilon ) (1-2 \epsilon )^2}{3 (1-3 \epsilon )} & \frac{(3-5 \epsilon ) (1-3 \epsilon ) (2-3 \epsilon )}{2 (1-4 \epsilon )} \\
             \frac{4 (3-5 \epsilon ) (1-2 \epsilon )^2}{3 (1-3 \epsilon )} & -\frac{(3-5 \epsilon ) (1-3 \epsilon ) (2-3 \epsilon )}{2 (1-4 \epsilon )}
     \end{pmatrix}\,,\\
    L_1&= f(\epsilon)^{-1}\begin{pmatrix}
        (3-5 \epsilon ) (1-2 \epsilon ) & 0 \\
        0 & \frac{(1-3 \epsilon ) \epsilon }{1-4 \epsilon }
    \end{pmatrix}\,.
\end{align}
The associator reads
\begin{equation}
    U_{01}=
    \begin{pmatrix}
        1+8 \zeta_3 \epsilon ^3-\frac{\pi ^4 \epsilon ^4}{9}+\ldots& -\frac{\pi ^2 \epsilon ^2}{3}-2 \zeta_3 \epsilon ^3-\frac{5 \pi ^4 \epsilon ^4}{18}+\ldots \\
        \frac{2 \pi ^2 \epsilon ^2}{3}-4 \zeta_3 \epsilon ^3+\frac{5 \pi ^4 \epsilon ^4}{9}+\ldots& 1-8 \zeta_3 \epsilon ^3-\frac{\pi ^4 \epsilon ^4}{9}+\ldots   
    \end{pmatrix}\,.
\end{equation}
At the point $x=1$ we have only ``naive'' limits, so $ [j_2]_{(1-x)^{-2\epsilon}}=0$. The constant $[j_1]_{(1-x)^0}$ reads
\begin{equation}
[j_1]_{(1-x)^0}=e^{4\epsilon\gamma_E}\frac{\Gamma (2-3 \epsilon ) \Gamma (1-\epsilon )^6 \Gamma (\epsilon )^2 \Gamma (4 \epsilon -2)}{\Gamma (4-5 \epsilon ) \Gamma (2-2 \epsilon )^2 \Gamma (2 \epsilon )}\,.
\end{equation}
It is easy to see that for $f(\epsilon)=\frac{1-3 \epsilon }{(1-5 \epsilon ) (2-5 \epsilon ) (1-4 \epsilon ) (1-2 \epsilon )^2 \epsilon ^2}$ the quantity
\begin{equation}
C_1=L_1{
    [j_1]_{(1-x)^0} \choose [j_2]_{(1-x)^{-2\epsilon}}}
\end{equation}
is uniformly transcendental. 
Then $C_0$ can be computed from $C_0=U_{01}C_1$ and is also  uniformly transcendental. On the other hand, we have
\begin{equation}
    C_0=L_0c_0=f(\epsilon)\begin{pmatrix}
        \frac{2 [j_1]_{x^0} (5 \epsilon -3) (2 \epsilon -1)^2}{3 (3 \epsilon -1)}+\frac{[j_1]_{x^{2-3\epsilon}} (3 \epsilon -2) (3 \epsilon -1) (5 \epsilon -3)}{2 (4 \epsilon -1)}
        \\
        \frac{4 [j_1]_{x^0} (2 \epsilon -1)^2 (5 \epsilon -3)}{3 (3 \epsilon -1)}-\frac{[j_1]_{x^{2-3\epsilon}} (3 \epsilon -2) (3 \epsilon -1) (5 \epsilon -3)}{2 (4 \epsilon -1)}
    \end{pmatrix}\,.
    \label{eq::C0}
\end{equation}
We see that each entry of $C_0$ is a linear combination of ``naive'' limit constant $[j_1]_{x^0} $, which corresponds to a specific on-shell vertex integral, and of the constant $[j_1]_{x^{2-3\epsilon}}$, which corresponds to a contribution of some non-trivial region in $x\to 0$ asymptotics. Thus the comparison to $C_0=U_{01}C_1$ does not allow for the extraction of $[j_1]_{x^0}$. However, if we consider $\tilde{C}_0=Q^{-1} C_0$, where  $Q=\small \begin{pmatrix}1 & -1 \\ 2 & 1 \end{pmatrix}$ is a transformation diagonalizing $S_0$ in Eq.~\eqref{eq:S0S1}, we obtain from Eq.~(\ref{eq::C0})
\begin{equation}
\tilde{C}_0=\begin{pmatrix}
    \frac{2 (2 \epsilon -1)^2 (5 \epsilon -3)}{3 (3 \epsilon -1)}  [j_1]_{x^0} \\
    -\frac{(3 \epsilon -2) (3 \epsilon -1) (5 \epsilon -3)}{2 (4 \epsilon -1)}[j_1]_{x^{2-3\epsilon}}
    \end{pmatrix}\,,
\end{equation}
and the comparison to $\tilde{C}_0=Q^{-1} U_{01}C_1$ immediately provides us a result for $[j_1]_{x^0}$.
The first entry of $\tilde{C}_0$ is expressed via on-shell vertex integral. Since $Q$ is rational numeric matrix, $\tilde{C}_0$ remains UT and we have achieved our goal.

In this way, we obtain UT bases for the vertex integrals with two massless legs through to weight 9, written in terms of multiple polylogarithms $G$ with argument 1 and indices $\{0,\pm1,$ $\pm i\sqrt{3},$ $e^{\pm i \pi/3},$ $e^{\pm 2 i \pi/3},$ $e^{\pm i \pi/3}/2 \}$. 
Employing the PSLQ algorithm~\cite{PSLQ}, these results can be expressed in terms of regular multiple zeta values.

We computed many integral coefficients in both approaches (direct integrations and differential equations), which allowed us to cross-check a substantial fraction of our results analytically.
To facilitate the checks of our results, we expressed \emph{all} master integrals in terms of finite integrals, which we define allowing also for higher dimensions and/or additional dots.
We determined all finite integrals to the required order in $\ep$ needed for complete weight 8 information, which occasionally involved also weight 9 contributions.
We also employed \texttt{Fiesta} \cite{smirnov2021fiesta5} for numerical checks of many integrals.
By performing these checks directly for finite integrals defined in $6-2\ep$ dimensions, we were able to achieve a typical relative agreement of $10^{-4}$ or better with modest run times.

For the master integrals entering the Sudakov form factor we obtain the following results through to weight 8:
{\footnotesize
\begin{align}\label{eq:intsolfirst}
I_{\text{p},1}^{(1)} &=  \pole{\frac{1}{\ep^8}} \Big(
   \mfrac{1}{576}
 \Big)+ \pole{\frac{1}{\ep^6}} \Big(
   \mfrac{17}{288}\zeta_2
 \Big)+ \pole{\frac{1}{\ep^5}} \Big(
   \mfrac{89}{432}\zeta_3
 \Big)+ \pole{\frac{1}{\ep^4}} \Big(
   \mfrac{677}{720}\zeta_2^2
 \Big)+ \pole{\frac{1}{\ep^3}} \Big(
   \mfrac{5489}{720}\zeta_5
  + \mfrac{487}{216}\zeta_3 \zeta_2
 \Big)+ \pole{\frac{1}{\ep^2}} \Big(
   \mfrac{1571}{324}\zeta_3^2 
 \notag\\ &\quad  + \mfrac{3919}{420}\zeta_2^3 
 \Big) + \pole{\frac{1}{\ep}} \Big(
   \mfrac{77677}{2016}\zeta_7
  + \mfrac{16543}{360}\zeta_5 \zeta_2
  + \mfrac{4957}{540}\zeta_3 \zeta_2^2
 \Big)+  \Big(
   \mfrac{727}{10}\zeta_{5,3}
  + \mfrac{15514}{135}\zeta_5 \zeta_3
  - \mfrac{4181}{162}\zeta_3^2 \zeta_2
  + \mfrac{232093}{126000}\zeta_2^4
 \Big) \notag\\ &\quad  + \order{\ep},\\
I_{\text{p},2}^{(2)} &=  \pole{\frac{1}{\ep^8}} \Big(
   \mfrac{1}{144}
 \Big)+ \pole{\frac{1}{\ep^6}} \Big(
  - \mfrac{13}{144}\zeta_2
 \Big)+ \pole{\frac{1}{\ep^5}} \Big(
  - \mfrac{577}{432}\zeta_3
 \Big)+ \pole{\frac{1}{\ep^4}} \Big(
  - \mfrac{269}{80}\zeta_2^2
 \Big)+ \pole{\frac{1}{\ep^3}} \Big(
  - \mfrac{4309}{720}\zeta_5
  - \mfrac{236}{27}\zeta_3 \zeta_2 
 \Big) \notag\\ &\quad + \pole{\frac{1}{\ep^2}} \Big(
   \mfrac{115529}{1296}\zeta_3^2
  - \mfrac{13721}{1260}\zeta_2^3
 \Big)+ \pole{\frac{1}{\ep}} \Big(
   \mfrac{958499}{1008}\zeta_7
  + \mfrac{6442}{45}\zeta_5 \zeta_2
  + \mfrac{16141}{120}\zeta_3 \zeta_2^2
 \Big)+  \Big(
  - \mfrac{4490}{3}\zeta_{5,3}
  + \mfrac{340829}{1080}\zeta_5 \zeta_3  \notag\\ &\quad
  + \mfrac{358051}{324}\zeta_3^2 \zeta_2
  + \mfrac{39385301}{25200}\zeta_2^4
 \Big) + \order{\ep},\\
I_{\text{p},3}^{(3)} &=  \pole{\frac{1}{\ep^8}} \Big(
  - \mfrac{1}{288}
 \Big)+ \pole{\frac{1}{\ep^6}} \Big(
  - \mfrac{17}{144}\zeta_2
 \Big)+ \pole{\frac{1}{\ep^5}} \Big(
  - \mfrac{233}{216}\zeta_3
 \Big)+ \pole{\frac{1}{\ep^4}} \Big(
  - \mfrac{173}{360}\zeta_2^2
 \Big)+ \pole{\frac{1}{\ep^3}} \Big(
  - \mfrac{16529}{360}\zeta_5
  + \mfrac{2033}{108}\zeta_3 \zeta_2
 \Big)\notag\\ &\quad + \pole{\frac{1}{\ep^2}} \Big(
  - \mfrac{8717}{162}\zeta_3^2
  - \mfrac{615}{14}\zeta_2^3
 \Big)+ \pole{\frac{1}{\ep}} \Big(
  - \mfrac{3335575}{1008}\zeta_7
  + \mfrac{31937}{180}\zeta_5 \zeta_2
  - \mfrac{30589}{270}\zeta_3 \zeta_2^2
 \Big)+  \Big(
   \mfrac{13891}{5}\zeta_{5,3}
  + \mfrac{26134}{135}\zeta_5 \zeta_3
  \notag\\ &\quad  - \mfrac{83065}{81}\zeta_3^2 \zeta_2
  - \mfrac{231920251}{63000}\zeta_2^4
 \Big) + \order{\ep},\\
I_{\text{p},4}^{(4)} &=  \pole{\frac{1}{\ep^8}} \Big(
   \mfrac{1}{288}
 \Big)+ \pole{\frac{1}{\ep^6}} \Big(
   \mfrac{17}{144}\zeta_2
 \Big)+ \pole{\frac{1}{\ep^5}} \Big(
   \mfrac{89}{216}\zeta_3
 \Big)+ \pole{\frac{1}{\ep^4}} \Big(
   \mfrac{533}{360}\zeta_2^2
 \Big)+ \pole{\frac{1}{\ep^3}} \Big(
   \mfrac{7469}{360}\zeta_5
  + \mfrac{163}{108}\zeta_3 \zeta_2
 \Big)+ \pole{\frac{1}{\ep^2}} \Big(
   \mfrac{1150}{81}\zeta_3^2
  + \mfrac{218}{35}\zeta_2^3
 \Big)\notag\\ &\quad  + \pole{\frac{1}{\ep}} \Big(
   \mfrac{124771}{252}\zeta_7
  - \mfrac{26657}{180}\zeta_5 \zeta_2
  - \mfrac{8507}{270}\zeta_3 \zeta_2^2
 \Big)+  \Big(
   \mint{33}\zeta_{5,3}
  + \mfrac{116303}{135}\zeta_5 \zeta_3
  - \mfrac{1022}{81}\zeta_3^2 \zeta_2
  + \mfrac{1180217}{3150}\zeta_2^4
 \Big) + \order{\ep},\\
I_{\text{p},5}^{(5)} &=  \pole{\frac{1}{\ep^8}} \Big(
   \mfrac{1}{72}
 \Big)+ \pole{\frac{1}{\ep^6}} \Big(
  - \mfrac{13}{72}\zeta_2
 \Big)+ \pole{\frac{1}{\ep^5}} \Big(
  - \mfrac{577}{216}\zeta_3
 \Big)+ \pole{\frac{1}{\ep^4}} \Big(
  - \mfrac{887}{120}\zeta_2^2
 \Big)+ \pole{\frac{1}{\ep^3}} \Big(
  - \mfrac{21109}{360}\zeta_5
  - \mfrac{4}{27}\zeta_3 \zeta_2
 \Big)\notag\\ &\quad  + \pole{\frac{1}{\ep^2}} \Big(
   \mfrac{193721}{648}\zeta_3^2
  - \mfrac{2897}{30}\zeta_2^3
 \Big)+ \pole{\frac{1}{\ep}} \Big(
  - \mfrac{239761}{504}\zeta_7
  - \mfrac{30136}{45}\zeta_5 \zeta_2
  + \mfrac{184207}{180}\zeta_3 \zeta_2^2
 \Big)+  \Big(
  - \mfrac{86152}{15}\zeta_{5,3}
  \notag\\ &\quad   + \mfrac{2197469}{540}\zeta_5 \zeta_3
  + \mfrac{48343}{162}\zeta_3^2 \zeta_2
  + \mfrac{113119649}{63000}\zeta_2^4
 \Big) + \order{\ep},\\
I_{\text{p},6}^{(6)} &=  \pole{\frac{1}{\ep^8}} \Big(
   \mfrac{1}{576}
 \Big)+ \pole{\frac{1}{\ep^6}} \Big(
   \mfrac{7}{144}\zeta_2
 \Big)+ \pole{\frac{1}{\ep^5}} \Big(
   \mfrac{169}{864}\zeta_3
 \Big)+ \pole{\frac{1}{\ep^4}} \Big(
   \mfrac{713}{1440}\zeta_2^2
 \Big)+ \pole{\frac{1}{\ep^3}} \Big(
   \mfrac{3013}{1440}\zeta_5
  + \mfrac{115}{216}\zeta_3 \zeta_2
 \Big)+ \pole{\frac{1}{\ep^2}} \Big(
  - \mfrac{13919}{2592}\zeta_3^2
  \notag\\ &\quad  + \mfrac{1759}{7560}\zeta_2^3
 \Big)+ \pole{\frac{1}{\ep}} \Big(
  - \mfrac{135691}{672}\zeta_7
  + \mfrac{23921}{360}\zeta_5 \zeta_2
  - \mfrac{38863}{2160}\zeta_3 \zeta_2^2
 \Big)+  \Big(
   \mfrac{3443}{180}\zeta_{5,3}
  - \mfrac{1103603}{2160}\zeta_5 \zeta_3
  + \mfrac{65419}{648}\zeta_3^2 \zeta_2
  \notag\\ &\quad  - \mfrac{32463187}{252000}\zeta_2^4
 \Big) + \order{\ep},\\
I_{\text{p},7}^{(7)} &=  \pole{\frac{1}{\ep^8}} \Big(
   \mfrac{11}{576}
 \Big)+ \pole{\frac{1}{\ep^6}} \Big(
   \mfrac{11}{48}\zeta_2
 \Big)+ \pole{\frac{1}{\ep^5}} \Big(
   \mfrac{1937}{864}\zeta_3
 \Big)+ \pole{\frac{1}{\ep^4}} \Big(
   \mfrac{487}{360}\zeta_2^2
 \Big)+ \pole{\frac{1}{\ep^3}} \Big(
   \mfrac{94313}{1440}\zeta_5
  - \mfrac{1505}{48}\zeta_3 \zeta_2
 \Big)+ \pole{\frac{1}{\ep^2}} \Big(
  - \mfrac{14483}{324}\zeta_3^2
  \notag\\ &\quad  + \mfrac{35053}{1260}\zeta_2^3
 \Big)+ \pole{\frac{1}{\ep}} \Big(
   \mfrac{6002449}{4032}\zeta_7
  - \mfrac{192539}{240}\zeta_5 \zeta_2
  - \mfrac{10093}{1080}\zeta_3 \zeta_2^2
 \Big)+  \Big(
   \mfrac{40379}{30}\zeta_{5,3}
  - \mfrac{191423}{270}\zeta_5 \zeta_3
  + \mfrac{20023}{54}\zeta_3^2 \zeta_2
 \notag\\ &\quad   - \mfrac{53988017}{84000}\zeta_2^4
 \Big) + \order{\ep},\\
I_{\text{p},8}^{(9)} &=  \pole{\frac{1}{\ep^8}} \Big(
   \mfrac{1}{576}
 \Big)+ \pole{\frac{1}{\ep^6}} \Big(
   \mfrac{1}{24}\zeta_2
 \Big)+ \pole{\frac{1}{\ep^5}} \Big(
   \mfrac{163}{864}\zeta_3
 \Big)+ \pole{\frac{1}{\ep^4}} \Big(
   \mfrac{161}{160}\zeta_2^2
 \Big)+ \pole{\frac{1}{\ep^3}} \Big(
   \mfrac{5803}{1440}\zeta_5
  + \mfrac{253}{36}\zeta_3 \zeta_2
 \Big)+ \pole{\frac{1}{\ep^2}} \Big(
   \mfrac{59509}{2592}\zeta_3^2
  + \mfrac{119}{6}\zeta_2^3
 \Big)\notag\\ &\quad  + \pole{\frac{1}{\ep}} \Big(
   \mfrac{2284607}{4032}\zeta_7
  - \mfrac{3449}{120}\zeta_5 \zeta_2
  + \mfrac{19081}{240}\zeta_3 \zeta_2^2
 \Big)+  \Big(
  - \mfrac{157}{15}\zeta_{5,3}
  + \mfrac{3485149}{2160}\zeta_5 \zeta_3
  - \mfrac{23813}{108}\zeta_3^2 \zeta_2
  + \mfrac{12188279}{28000}\zeta_2^4
 \Big)  \notag\\ &\quad  + \order{\ep},\\
I_{\text{p},9}^{(10)} &=  \pole{\frac{1}{\ep^8}} \Big(
  - \mfrac{13}{576}
 \Big)+ \pole{\frac{1}{\ep^6}} \Big(
   \mfrac{5}{48}\zeta_2
 \Big)+ \pole{\frac{1}{\ep^5}} \Big(
   \mfrac{743}{864}\zeta_3
 \Big)+ \pole{\frac{1}{\ep^4}} \Big(
   \mfrac{167}{480}\zeta_2^2
 \Big)+ \pole{\frac{1}{\ep^3}} \Big(
   \mfrac{82931}{1440}\zeta_5
  - \mfrac{179}{9}\zeta_3 \zeta_2
 \Big)+ \pole{\frac{1}{\ep^2}} \Big(
   \mfrac{425345}{2592}\zeta_3^2
  \notag\\ &\quad  + \mfrac{4163}{90}\zeta_2^3
 \Big)+ \pole{\frac{1}{\ep}} \Big(
   \mfrac{16246723}{4032}\zeta_7
  - \mfrac{1499}{4}\zeta_5 \zeta_2
  + \mfrac{313829}{720}\zeta_3 \zeta_2^2
 \Big)+  \Big(
  - \mfrac{176657}{30}\zeta_{5,3}
  - \mfrac{10377031}{2160}\zeta_5 \zeta_3
  + \mfrac{68821}{27}\zeta_3^2 \zeta_2
  \notag\\ &\quad  + \mfrac{443297431}{84000}\zeta_2^4
 \Big) + \order{\ep},\\
I_{\text{p},10}^{(12)} &=  \pole{\frac{1}{\ep^8}} \Big(
  - \mfrac{1}{72}
 \Big)+ \pole{\frac{1}{\ep^6}} \Big(
  - \mfrac{29}{144}\zeta_2
 \Big)+ \pole{\frac{1}{\ep^5}} \Big(
  - \mfrac{577}{432}\zeta_3
 \Big)+ \pole{\frac{1}{\ep^4}} \Big(
   \mfrac{31}{240}\zeta_2^2
 \Big)+ \pole{\frac{1}{\ep^3}} \Big(
  - \mfrac{36367}{720}\zeta_5
  + \mfrac{4019}{108}\zeta_3 \zeta_2
 \Big)\notag\\ &\quad  + \pole{\frac{1}{\ep^2}} \Big(
   \mfrac{128729}{1296}\zeta_3^2
  - \mfrac{4741}{252}\zeta_2^3
 \Big)+ \pole{\frac{1}{\ep}} \Big(
  - \mfrac{7604257}{4032}\zeta_7
  + \mfrac{361843}{360}\zeta_5 \zeta_2
  + \mfrac{21493}{360}\zeta_3 \zeta_2^2
 \Big)+  \Big(
   \mfrac{3205}{12}\zeta_{5,3}
  + \mfrac{6666179}{1080}\zeta_5 \zeta_3
  \notag\\ &\quad  - \mfrac{797531}{648}\zeta_3^2 \zeta_2
  - \mfrac{1336955}{2016}\zeta_2^4
 \Big) + \order{\ep},\\
I_{\text{p},11}^{(12)} &=  \pole{\frac{1}{\ep^8}} \Big(
   \mfrac{1}{144}
 \Big)+ \pole{\frac{1}{\ep^6}} \Big(
  - \mfrac{1}{144}\zeta_2
 \Big)+ \pole{\frac{1}{\ep^5}} \Big(
  - \mfrac{1}{432}\zeta_3
 \Big)+ \pole{\frac{1}{\ep^4}} \Big(
   \mfrac{57}{80}\zeta_2^2
 \Big)+ \pole{\frac{1}{\ep^3}} \Big(
  - \mfrac{38149}{720}\zeta_5
  + \mfrac{1061}{54}\zeta_3 \zeta_2
 \Big)\notag\\ &\quad  + \pole{\frac{1}{\ep^2}} \Big(
  - \mfrac{237775}{1296}\zeta_3^2
  - \mfrac{37363}{2520}\zeta_2^3
 \Big)+ \pole{\frac{1}{\ep}} \Big(
  - \mfrac{2582855}{504}\zeta_7
  + \mfrac{99319}{45}\zeta_5 \zeta_2
  - \mfrac{24527}{40}\zeta_3 \zeta_2^2
 \Big)+  \Big(
   \mfrac{204143}{30}\zeta_{5,3}
  \notag\\ &\quad  + \mfrac{694817}{1080}\zeta_5 \zeta_3
  + \mfrac{316049}{162}\zeta_3^2 \zeta_2
  - \mfrac{387942419}{84000}\zeta_2^4
 \Big) + \order{\ep},\\
I_{\text{p},12}^{(13)} &=  \pole{\frac{1}{\ep^8}} \Big(
   \mfrac{1}{576}
 \Big)+ \pole{\frac{1}{\ep^6}} \Big(
   \mfrac{1}{24}\zeta_2
 \Big)+ \pole{\frac{1}{\ep^5}} \Big(
   \mfrac{181}{864}\zeta_3
 \Big)+ \pole{\frac{1}{\ep^4}} \Big(
   \mfrac{57}{80}\zeta_2^2
 \Big)+ \pole{\frac{1}{\ep^3}} \Big(
   \mfrac{5833}{1440}\zeta_5
  + \mfrac{595}{144}\zeta_3 \zeta_2
 \Big)+ \pole{\frac{1}{\ep^2}} \Big(
   \mfrac{2083}{162}\zeta_3^2
  + \mfrac{33163}{2520}\zeta_2^3
 \Big)\notag\\ &\quad  + \pole{\frac{1}{\ep}} \Big(
   \mfrac{1623313}{8064}\zeta_7
  + \mfrac{7001}{120}\zeta_5 \zeta_2
  + \mfrac{4063}{80}\zeta_3 \zeta_2^2
 \Big)+  \Big(
  - \mfrac{1567}{120}\zeta_{5,3}
  + \mfrac{201041}{270}\zeta_5 \zeta_3
  - \mfrac{1513}{216}\zeta_3^2 \zeta_2
  + \mfrac{3031177}{12000}\zeta_2^4
 \Big)\notag\\ &\quad  + \order{\ep},\\
I_{\text{p},13}^{(14)} &=  \pole{\frac{1}{\ep^8}} \Big(
   \mfrac{23}{576}
 \Big)+ \pole{\frac{1}{\ep^6}} \Big(
  - \mfrac{47}{144}\zeta_2
 \Big)+ \pole{\frac{1}{\ep^5}} \Big(
  - \mfrac{1789}{864}\zeta_3
 \Big)+ \pole{\frac{1}{\ep^4}} \Big(
  - \mfrac{433}{288}\zeta_2^2
 \Big)+ \pole{\frac{1}{\ep^3}} \Big(
  - \mfrac{60961}{1440}\zeta_5
  + \mfrac{4765}{216}\zeta_3 \zeta_2
 \Big)\notag\\ &\quad  + \pole{\frac{1}{\ep^2}} \Big(
   \mfrac{134567}{2592}\zeta_3^2
  - \mfrac{52957}{2520}\zeta_2^3
 \Big)+ \pole{\frac{1}{\ep}} \Big(
  - \mfrac{2434597}{2016}\zeta_7
  + \mfrac{97279}{360}\zeta_5 \zeta_2
  + \mfrac{295021}{2160}\zeta_3 \zeta_2^2
 \Big)+  \Big(
  - \mfrac{7319}{60}\zeta_{5,3}
  \notag\\ &\quad  + \mfrac{2126603}{2160}\zeta_5 \zeta_3
  - \mfrac{60533}{648}\zeta_3^2 \zeta_2
  - \mfrac{14105297}{42000}\zeta_2^4
 \Big) + \order{\ep},\\
I_{\text{p},14}^{(17)} &=  \pole{\frac{1}{\ep^8}} \Big(
  - \mfrac{3}{64}
 \Big)+ \pole{\frac{1}{\ep^6}} \Big(
   \mfrac{31}{96}\zeta_2
 \Big)+ \pole{\frac{1}{\ep^5}} \Big(
   \mfrac{3}{4}\zeta_3
 \Big)+ \pole{\frac{1}{\ep^4}} \Big(
  - \mfrac{6541}{1440}\zeta_2^2
 \Big)+ \pole{\frac{1}{\ep^3}} \Big(
  - \mfrac{1063}{20}\zeta_5
  - \mfrac{781}{36}\zeta_3 \zeta_2
 \Big)+ \pole{\frac{1}{\ep^2}} \Big(
   \mfrac{2741}{144}\zeta_3^2
  \notag\\ &\quad  - \mfrac{192937}{2016}\zeta_2^3
 \Big)+ \pole{\frac{1}{\ep}} \Big(
  - \mfrac{3518471}{2688}\zeta_7
  - \mfrac{32669}{80}\zeta_5 \zeta_2
  + \mfrac{719429}{2160}\zeta_3 \zeta_2^2
 \Big)+  \Big(
   \mfrac{4559}{10}\zeta_{5,3}
  + \mfrac{268141}{40}\zeta_5 \zeta_3
  + \mfrac{10627}{144}\zeta_3^2 \zeta_2
  \notag\\ &\quad  - \mfrac{55582357}{48000}\zeta_2^4
 \Big) + \order{\ep},\\
I_{\text{p},15}^{(17)} &=  \pole{\frac{1}{\ep^8}} \Big(
   \mfrac{1}{576}
 \Big)+ \pole{\frac{1}{\ep^6}} \Big(
  - \mfrac{1}{8}\zeta_2
 \Big)+ \pole{\frac{1}{\ep^5}} \Big(
  - \mfrac{319}{432}\zeta_3
 \Big)+ \pole{\frac{1}{\ep^4}} \Big(
   \mfrac{1201}{2880}\zeta_2^2
 \Big)+ \pole{\frac{1}{\ep^3}} \Big(
  - \mfrac{1373}{360}\zeta_5
  + \mfrac{2353}{144}\zeta_3 \zeta_2
 \Big)+ \pole{\frac{1}{\ep^2}} \Big(
   \mfrac{7328}{81}\zeta_3^2
  \notag\\ &\quad  + \mfrac{67729}{4032}\zeta_2^3
 \Big)+ \pole{\frac{1}{\ep}} \Big(
   \mfrac{3392399}{4032}\zeta_7
  - \mfrac{27833}{120}\zeta_5 \zeta_2
  + \mfrac{890263}{4320}\zeta_3 \zeta_2^2
 \Big)+  \Big(
  - \mfrac{7591}{40}\zeta_{5,3}
  + \mfrac{3939533}{1080}\zeta_5 \zeta_3
  - \mfrac{35209}{27}\zeta_3^2 \zeta_2
  \notag\\ &\quad  + \mfrac{952297279}{2016000}\zeta_2^4
 \Big) + \order{\ep},\\
I_{\text{p},16}^{(19)} &=  \pole{\frac{1}{\ep^8}} \Big(
  - \mfrac{1}{96}
 \Big)+ \pole{\frac{1}{\ep^6}} \Big(
  - \mfrac{53}{288}\zeta_2
 \Big)+ \pole{\frac{1}{\ep^5}} \Big(
  - \mfrac{337}{288}\zeta_3
 \Big)+ \pole{\frac{1}{\ep^4}} \Big(
  - \mfrac{637}{240}\zeta_2^2
 \Big)+ \pole{\frac{1}{\ep^3}} \Big(
  - \mfrac{27601}{480}\zeta_5
  + \mfrac{1541}{216}\zeta_3 \zeta_2
 \Big)\notag\\ &\quad  + \pole{\frac{1}{\ep^2}} \Big(
  - \mfrac{4069}{288}\zeta_3^2
  - \mfrac{524371}{10080}\zeta_2^3
 \Big)+ \pole{\frac{1}{\ep}} \Big(
  - \mfrac{1698595}{672}\zeta_7
  + \mfrac{33556}{45}\zeta_5 \zeta_2
  + \mfrac{8807}{360}\zeta_3 \zeta_2^2
 \Big)+  \Big(
   \mfrac{4893}{10}\zeta_{5,3}
  - \mfrac{83399}{80}\zeta_5 \zeta_3
  \notag\\ &\quad  + \mfrac{924893}{648}\zeta_3^2 \zeta_2
  - \mfrac{1110914569}{1008000}\zeta_2^4
 \Big) + \order{\ep},\\
I_{\text{p},17}^{(19)} &=  \pole{\frac{1}{\ep^8}} \Big(
   \mfrac{5}{288}
 \Big)+ \pole{\frac{1}{\ep^6}} \Big(
  - \mfrac{139}{288}\zeta_2
 \Big)+ \pole{\frac{1}{\ep^5}} \Big(
  - \mfrac{3809}{864}\zeta_3
 \Big)+ \pole{\frac{1}{\ep^4}} \Big(
  - \mfrac{2609}{240}\zeta_2^2
 \Big)+ \pole{\frac{1}{\ep^3}} \Big(
  - \mfrac{56425}{288}\zeta_5
  + \mfrac{10235}{432}\zeta_3 \zeta_2
 \Big)\notag\\ &\quad  + \pole{\frac{1}{\ep^2}} \Big(
   \mfrac{80651}{1296}\zeta_3^2
  - \mfrac{184073}{1008}\zeta_2^3
 \Big)+ \pole{\frac{1}{\ep}} \Big(
  - \mfrac{27831739}{4032}\zeta_7
  + \mfrac{630593}{720}\zeta_5 \zeta_2
  + \mfrac{39206}{45}\zeta_3 \zeta_2^2
 \Big)+  \Big(
   \mfrac{50049}{20}\zeta_{5,3}
  \notag\\ &\quad  + \mfrac{11313389}{1080}\zeta_5 \zeta_3
  + \mfrac{474319}{162}\zeta_3^2 \zeta_2
  - \mfrac{1865882993}{504000}\zeta_2^4
 \Big) + \order{\ep},\\
I_{\text{p},18}^{(21)} &=  \pole{\frac{1}{\ep^8}} \Big(
   \mfrac{1}{576}
 \Big)+ \pole{\frac{1}{\ep^6}} \Big(
   \mfrac{1}{36}\zeta_2
 \Big)+ \pole{\frac{1}{\ep^5}} \Big(
   \mfrac{151}{864}\zeta_3
 \Big)+ \pole{\frac{1}{\ep^4}} \Big(
   \mfrac{173}{288}\zeta_2^2
 \Big)+ \pole{\frac{1}{\ep^3}} \Big(
   \mfrac{5503}{1440}\zeta_5
  + \mfrac{505}{216}\zeta_3 \zeta_2
 \Big)+ \pole{\frac{1}{\ep^2}} \Big(
   \mfrac{9895}{2592}\zeta_3^2
  + \mfrac{6317}{720}\zeta_2^3
 \Big)\notag\\ &\quad  + \pole{\frac{1}{\ep}} \Big(
  - \mfrac{169789}{4032}\zeta_7
  + \mfrac{3419}{45}\zeta_5 \zeta_2
  + \mfrac{89593}{2160}\zeta_3 \zeta_2^2
 \Big)+  \Big(
   \mfrac{407}{15}\zeta_{5,3}
  - \mfrac{263897}{2160}\zeta_5 \zeta_3
  + \mfrac{41719}{162}\zeta_3^2 \zeta_2
  + \mfrac{43695623}{504000}\zeta_2^4
 \Big) \notag\\ &\quad + \order{\ep},\\
I_{\text{p},19}^{(25)} &=  \pole{\frac{1}{\ep^8}} \Big(
   \mfrac{1}{288}
 \Big)+ \pole{\frac{1}{\ep^6}} \Big(
   \mfrac{1}{144}\zeta_2
 \Big)+ \pole{\frac{1}{\ep^5}} \Big(
   \mfrac{209}{216}\zeta_3
 \Big)+ \pole{\frac{1}{\ep^4}} \Big(
   \mfrac{623}{120}\zeta_2^2
 \Big)+ \pole{\frac{1}{\ep^3}} \Big(
   \mfrac{39449}{360}\zeta_5
  - \mfrac{205}{108}\zeta_3 \zeta_2
 \Big)+ \pole{\frac{1}{\ep^2}} \Big(
   \mfrac{11621}{162}\zeta_3^2
  \notag\\ &\quad  + \mfrac{38501}{315}\zeta_2^3
 \Big)+ \pole{\frac{1}{\ep}} \Big(
   \mfrac{2997077}{504}\zeta_7
  - \mfrac{290821}{180}\zeta_5 \zeta_2
  + \mfrac{8023}{90}\zeta_3 \zeta_2^2
 \Big)+  \Big(
  - \mfrac{62426}{15}\zeta_{5,3}
  - \mfrac{493027}{135}\zeta_5 \zeta_3
  - \mfrac{210472}{81}\zeta_3^2 \zeta_2
  \notag\\ &\quad  + \mfrac{69281143}{15750}\zeta_2^4
 \Big) + \order{\ep},\\
I_{\text{p},20}^{(30)} &=  \pole{\frac{1}{\ep^8}} \Big(
   \mfrac{1}{288}
 \Big)+ \pole{\frac{1}{\ep^6}} \Big(
  - \mfrac{1}{32}\zeta_2
 \Big)+ \pole{\frac{1}{\ep^5}} \Big(
  - \mfrac{187}{864}\zeta_3
 \Big)+ \pole{\frac{1}{\ep^4}} \Big(
  - \mfrac{403}{720}\zeta_2^2
 \Big)+ \pole{\frac{1}{\ep^3}} \Big(
  - \mfrac{38659}{1440}\zeta_5
  + \mfrac{191}{36}\zeta_3 \zeta_2
 \Big)\notag\\ &\quad  + \pole{\frac{1}{\ep^2}} \Big(
  - \mfrac{14047}{2592}\zeta_3^2
  - \mfrac{284189}{10080}\zeta_2^3
 \Big)+ \pole{\frac{1}{\ep}} \Big(
  - \mfrac{1150361}{1008}\zeta_7
  + \mfrac{25019}{60}\zeta_5 \zeta_2
  - \mfrac{77089}{1080}\zeta_3 \zeta_2^2
 \Big)+  \Big(
  - \mfrac{7253}{15}\zeta_{5,3}
  \notag\\ &\quad  - \mfrac{2593651}{2160}\zeta_5 \zeta_3
  + \mfrac{40867}{108}\zeta_3^2 \zeta_2
  - \mfrac{7941559}{48000}\zeta_2^4
 \Big) + \order{\ep},\\
I_{\text{p},21}^{(13)} &=  \pole{\frac{1}{\ep^5}} \Big(
   \mfrac{1}{24}\zeta_3
 \Big)+ \pole{\frac{1}{\ep^3}} \Big(
   \mfrac{7}{12}\zeta_5
  - \mfrac{5}{12}\zeta_3 \zeta_2
 \Big)+ \pole{\frac{1}{\ep^2}} \Big(
  - \mfrac{193}{72}\zeta_3^2
  + \mfrac{6389}{2520}\zeta_2^3
 \Big)+ \pole{\frac{1}{\ep}} \Big(
   \mfrac{44}{3}\zeta_7
  + \mfrac{189}{4}\zeta_5 \zeta_2
  - \mfrac{2431}{120}\zeta_3 \zeta_2^2
 \Big)\notag\\ &\quad  +  \Big(
  - \mfrac{2621}{60}\zeta_{5,3}
  - \mfrac{45341}{180}\zeta_5 \zeta_3
  - \mfrac{9161}{72}\zeta_3^2 \zeta_2
  + \mfrac{228869}{5250}\zeta_2^4
 \Big) + \order{\ep},\\
I_{\text{p},22}^{(14)} &=  \pole{\frac{1}{\ep^6}} \Big(
   \mfrac{1}{48}\zeta_2
 \Big)+ \pole{\frac{1}{\ep^5}} \Big(
  - \mfrac{7}{48}\zeta_3
 \Big)+ \pole{\frac{1}{\ep^4}} \Big(
  - \mfrac{13}{240}\zeta_2^2
 \Big)+ \pole{\frac{1}{\ep^3}} \Big(
  - \mfrac{281}{48}\zeta_5
  + \mfrac{17}{9}\zeta_3 \zeta_2
 \Big)+ \pole{\frac{1}{\ep^2}} \Big(
   \mfrac{439}{144}\zeta_3^2
  - \mfrac{8053}{2520}\zeta_2^3
 \Big)\notag\\ &\quad  + \pole{\frac{1}{\ep}} \Big(
  - \mfrac{16135}{96}\zeta_7
  - \mfrac{544}{15}\zeta_5 \zeta_2
  + \mfrac{8497}{180}\zeta_3 \zeta_2^2
 \Big)+  \Big(
  - \mfrac{15557}{60}\zeta_{5,3}
  - \mfrac{3373}{360}\zeta_5 \zeta_3
  - \mfrac{55691}{216}\zeta_3^2 \zeta_2
  + \mfrac{3474517}{126000}\zeta_2^4
 \Big) \notag\\ &\quad + \order{\ep},\\
I_{\text{p},23}^{(14)} &=  \pole{\frac{1}{\ep^4}} \Big(
  - \mfrac{7}{20}\zeta_2^2
 \Big)+ \pole{\frac{1}{\ep^3}} \Big(
  - \mfrac{377}{24}\zeta_5
  + \mfrac{97}{12}\zeta_3 \zeta_2
 \Big)+ \pole{\frac{1}{\ep^2}} \Big(
   \mfrac{433}{24}\zeta_3^2
  - \mfrac{8531}{840}\zeta_2^3
 \Big)+ \pole{\frac{1}{\ep}} \Big(
  - \mfrac{92183}{64}\zeta_7
  + \mfrac{1387}{3}\zeta_5 \zeta_2
  \notag\\ &\quad  + \mfrac{5609}{60}\zeta_3 \zeta_2^2
 \Big)+  \Big(
   \mfrac{30512}{15}\zeta_{5,3}
  + \mfrac{168463}{36}\zeta_5 \zeta_3
  - \mfrac{17131}{36}\zeta_3^2 \zeta_2
  - \mfrac{32238699}{28000}\zeta_2^4
 \Big) + \order{\ep},\\
I_{\text{p},24}^{(14)} &=  \pole{\frac{1}{\ep^8}} \Big(
   \mfrac{5}{48}
 \Big)+ \pole{\frac{1}{\ep^6}} \Big(
  - \mfrac{65}{72}\zeta_2
 \Big)+ \pole{\frac{1}{\ep^5}} \Big(
  - \mfrac{293}{48}\zeta_3
 \Big)+ \pole{\frac{1}{\ep^4}} \Big(
  - \mfrac{2171}{480}\zeta_2^2
 \Big)+ \pole{\frac{1}{\ep^3}} \Big(
  - \mfrac{4019}{48}\zeta_5
  + \mfrac{11495}{216}\zeta_3 \zeta_2
 \Big)\notag\\ &\quad  + \pole{\frac{1}{\ep^2}} \Big(
   \mfrac{82361}{432}\zeta_3^2
  - \mfrac{163871}{10080}\zeta_2^3
 \Big)+ \pole{\frac{1}{\ep}} \Big(
  - \mfrac{707127}{448}\zeta_7
  + \mfrac{41611}{72}\zeta_5 \zeta_2
  + \mfrac{47639}{144}\zeta_3 \zeta_2^2
 \Big)+  \Big(
   \mfrac{42559}{60}\zeta_{5,3}
  + \mfrac{1714063}{360}\zeta_5 \zeta_3
  \notag\\ &\quad  - \mfrac{967087}{648}\zeta_3^2 \zeta_2
  - \mfrac{102398431}{144000}\zeta_2^4
 \Big) + \order{\ep},\\
I_{\text{p},25}^{(17)} &=  \pole{\frac{1}{\ep^8}} \Big(
   \mfrac{1}{96}
 \Big)+ \pole{\frac{1}{\ep^6}} \Big(
  - \mfrac{11}{96}\zeta_2
 \Big)+ \pole{\frac{1}{\ep^5}} \Big(
  - \mfrac{11}{9}\zeta_3
 \Big)+ \pole{\frac{1}{\ep^4}} \Big(
  - \mfrac{2743}{960}\zeta_2^2
 \Big)+ \pole{\frac{1}{\ep^3}} \Big(
  - \mfrac{2329}{80}\zeta_5
  - \mfrac{11}{36}\zeta_3 \zeta_2
 \Big)+ \pole{\frac{1}{\ep^2}} \Big(
   \mfrac{26141}{864}\zeta_3^2
  \notag\\ &\quad  - \mfrac{642007}{20160}\zeta_2^3
 \Big)+ \pole{\frac{1}{\ep}} \Big(
  - \mfrac{1937119}{2688}\zeta_7
  + \mfrac{1101}{20}\zeta_5 \zeta_2
  + \mfrac{169157}{1440}\zeta_3 \zeta_2^2
 \Big)+  \Big(
   \mfrac{9449}{120}\zeta_{5,3}
  + \mfrac{399373}{240}\zeta_5 \zeta_3
  + \mfrac{27517}{432}\zeta_3^2 \zeta_2
  \notag\\ &\quad  - \mfrac{754079597}{2016000}\zeta_2^4
 \Big) + \order{\ep},\\
I_{\text{p},26}^{(17)} &=  \pole{\frac{1}{\ep^8}} \Big(
   \mfrac{5}{144}
 \Big)+ \pole{\frac{1}{\ep^6}} \Big(
  - \mfrac{2}{9}\zeta_2
 \Big)+ \pole{\frac{1}{\ep^5}} \Big(
  - \mfrac{331}{216}\zeta_3
 \Big)+ \pole{\frac{1}{\ep^4}} \Big(
   \mfrac{1171}{240}\zeta_2^2
 \Big)+ \pole{\frac{1}{\ep^3}} \Big(
   \mfrac{3857}{36}\zeta_5
  + \mfrac{2041}{216}\zeta_3 \zeta_2
 \Big)+ \pole{\frac{1}{\ep^2}} \Big(
   \mfrac{72223}{1296}\zeta_3^2
  \notag\\ &\quad  + \mfrac{67171}{504}\zeta_2^3
 \Big)+ \pole{\frac{1}{\ep}} \Big(
   \mfrac{14485195}{4032}\zeta_7
  - \mfrac{21157}{45}\zeta_5 \zeta_2
  - \mfrac{39241}{180}\zeta_3 \zeta_2^2
 \Big)+  \Big(
   \mfrac{1664}{5}\zeta_{5,3}
  - \mfrac{4344341}{1080}\zeta_5 \zeta_3
  - \mfrac{343849}{324}\zeta_3^2 \zeta_2
  \notag\\ &\quad  + \mfrac{339481019}{252000}\zeta_2^4
 \Big) + \order{\ep},\\
I_{\text{p},27}^{(17)} &=  \pole{\frac{1}{\ep^5}} \Big(
   \mfrac{1}{48}\zeta_3
 \Big)+ \pole{\frac{1}{\ep^4}} \Big(
  - \mfrac{7}{160}\zeta_2^2
 \Big)+ \pole{\frac{1}{\ep^3}} \Big(
   \mfrac{91}{48}\zeta_5
  - \mfrac{113}{48}\zeta_3 \zeta_2
 \Big)+ \pole{\frac{1}{\ep^2}} \Big(
   \mfrac{1063}{288}\zeta_3^2
  - \mfrac{13751}{5040}\zeta_2^3
 \Big)+ \pole{\frac{1}{\ep}} \Big(
   \mfrac{46099}{192}\zeta_7
  \notag\\ &\quad  - \mfrac{9517}{48}\zeta_5 \zeta_2
  + \mfrac{3617}{80}\zeta_3 \zeta_2^2
 \Big)+  \Big(
   \mfrac{6849}{20}\zeta_{5,3}
  + \mfrac{999253}{720}\zeta_5 \zeta_3
  - \mfrac{943}{36}\zeta_3^2 \zeta_2
  - \mfrac{11407999}{72000}\zeta_2^4
 \Big) + \order{\ep},\\
I_{\text{p},28}^{(17)} &=  \pole{\frac{1}{\ep^6}} \Big(
   \mfrac{1}{24}\zeta_2
 \Big)+ \pole{\frac{1}{\ep^5}} \Big(
  - \mfrac{1}{8}\zeta_3
 \Big)+ \pole{\frac{1}{\ep^4}} \Big(
   \mfrac{1}{60}\zeta_2^2
 \Big)+ \pole{\frac{1}{\ep^3}} \Big(
   \mfrac{20}{3}\zeta_5
  - \mfrac{29}{36}\zeta_3 \zeta_2
 \Big)+ \pole{\frac{1}{\ep^2}} \Big(
   \mfrac{703}{24}\zeta_3^2
  + \mfrac{11281}{1260}\zeta_2^3
 \Big)+ \pole{\frac{1}{\ep}} \Big(
   \mfrac{25855}{48}\zeta_7
  \notag\\ &\quad  - \mfrac{121}{5}\zeta_5 \zeta_2
  + \mfrac{2807}{45}\zeta_3 \zeta_2^2
 \Big)+  \Big(
  - \mfrac{12944}{15}\zeta_{5,3}
  - \mfrac{37447}{180}\zeta_5 \zeta_3
  - \mfrac{7802}{27}\zeta_3^2 \zeta_2
  + \mfrac{103912481}{126000}\zeta_2^4
 \Big) + \order{\ep},\\
I_{\text{p},29}^{(19)} &=  \pole{\frac{1}{\ep^8}} \Big(
  - \mfrac{1}{36}
 \Big)+ \pole{\frac{1}{\ep^6}} \Big(
  - \mfrac{1}{16}\zeta_2
 \Big)+ \pole{\frac{1}{\ep^5}} \Big(
  - \mfrac{107}{432}\zeta_3
 \Big)+ \pole{\frac{1}{\ep^4}} \Big(
   \mfrac{9}{2}\zeta_2^2
 \Big)+ \pole{\frac{1}{\ep^3}} \Big(
   \mfrac{18091}{720}\zeta_5
  + \mfrac{467}{12}\zeta_3 \zeta_2
 \Big)+ \pole{\frac{1}{\ep^2}} \Big(
   \mfrac{155179}{1296}\zeta_3^2
  \notag\\ &\quad  + \mfrac{348347}{5040}\zeta_2^3
 \Big)+ \pole{\frac{1}{\ep}} \Big(
   \mfrac{920383}{1008}\zeta_7
  + \mfrac{25153}{60}\zeta_5 \zeta_2
  + \mfrac{1225}{24}\zeta_3 \zeta_2^2
 \Big)+  \Big(
   \mfrac{1189}{6}\zeta_{5,3}
  + \mfrac{3745369}{1080}\zeta_5 \zeta_3
  - \mfrac{40175}{24}\zeta_3^2 \zeta_2
  \notag\\ &\quad  + \mfrac{12545753}{14400}\zeta_2^4
 \Big) + \order{\ep},\\
I_{\text{p},30}^{(19)} &=  \pole{\frac{1}{\ep^5}} \Big(
  - \mfrac{7}{24}\zeta_3
 \Big)+ \pole{\frac{1}{\ep^4}} \Big(
  - \mfrac{5}{48}\zeta_2^2
 \Big)+ \pole{\frac{1}{\ep^3}} \Big(
   \mfrac{69}{8}\zeta_5
  + \mfrac{7}{4}\zeta_3 \zeta_2
 \Big)+ \pole{\frac{1}{\ep^2}} \Big(
   \mfrac{1885}{72}\zeta_3^2
  + \mfrac{8131}{1008}\zeta_2^3
 \Big)+ \pole{\frac{1}{\ep}} \Big(
   \mfrac{15071}{48}\zeta_7
  - \mfrac{266}{3}\zeta_5 \zeta_2
  \notag\\ &\quad  + \mfrac{16751}{360}\zeta_3 \zeta_2^2
 \Big)+  \Big(
   \mfrac{3269}{30}\zeta_{5,3}
  + \mfrac{11363}{20}\zeta_5 \zeta_3
  - \mfrac{641}{3}\zeta_3^2 \zeta_2
  + \mfrac{32252011}{168000}\zeta_2^4
 \Big) + \order{\ep},\\
I_{\text{p},31}^{(19)} &=  \pole{\frac{1}{\ep^8}} \Big(
  - \mfrac{11}{288}
 \Big)+ \pole{\frac{1}{\ep^6}} \Big(
   \mfrac{65}{288}\zeta_2
 \Big)+ \pole{\frac{1}{\ep^5}} \Big(
  - \mfrac{2005}{864}\zeta_3
 \Big)+ \pole{\frac{1}{\ep^4}} \Big(
   \mfrac{63}{80}\zeta_2^2
 \Big)+ \pole{\frac{1}{\ep^3}} \Big(
   \mfrac{115559}{1440}\zeta_5
  - \mfrac{4519}{432}\zeta_3 \zeta_2
 \Big)+ \pole{\frac{1}{\ep^2}} \Big(
   \mfrac{18203}{162}\zeta_3^2
  \notag\\ &\quad  + \mfrac{22915}{252}\zeta_2^3
 \Big)+ \pole{\frac{1}{\ep}} \Big(
   \mfrac{21937549}{4032}\zeta_7
  - \mfrac{125525}{144}\zeta_5 \zeta_2
  - \mfrac{4009}{20}\zeta_3 \zeta_2^2
 \Big)+  \Big(
  - \mfrac{287479}{60}\zeta_{5,3}
  - \mfrac{941249}{108}\zeta_5 \zeta_3
  - \mfrac{956701}{324}\zeta_3^2 \zeta_2
  \notag\\ &\quad  + \mfrac{183065759}{36000}\zeta_2^4
 \Big) + \order{\ep},\\
I_{\text{p},32}^{(30)} &=  \pole{\frac{1}{\ep^8}} \Big(
  - \mfrac{1}{12}
 \Big)+ \pole{\frac{1}{\ep^6}} \Big(
   \mfrac{35}{48}\zeta_2
 \Big)+ \pole{\frac{1}{\ep^5}} \Big(
   \mfrac{445}{144}\zeta_3
 \Big)+ \pole{\frac{1}{\ep^4}} \Big(
  - \mfrac{269}{240}\zeta_2^2
 \Big)+ \pole{\frac{1}{\ep^3}} \Big(
   \mfrac{2767}{80}\zeta_5
  - \mfrac{1433}{36}\zeta_3 \zeta_2
 \Big)+ \pole{\frac{1}{\ep^2}} \Big(
  - \mfrac{14051}{432}\zeta_3^2
  \notag\\ &\quad  - \mfrac{8363}{630}\zeta_2^3
 \Big)+ \pole{\frac{1}{\ep}} \Big(
   \mfrac{89105}{336}\zeta_7
  - \mfrac{3639}{4}\zeta_5 \zeta_2
  + \mfrac{162007}{360}\zeta_3 \zeta_2^2
 \Big)+  \Big(
   \mint{899}\zeta_{5,3}
  + \mfrac{25969}{8}\zeta_5 \zeta_3
  + \mfrac{31201}{27}\zeta_3^2 \zeta_2
  \notag\\ &\quad - \mfrac{2959007}{8400}\zeta_2^4
 \Big)  + \order{\ep}, \\
I_{1}^{(21)} &= I_{\mathrm{p}, 18}^{(21)} ,\\
I_{2}^{(22)} &=  \pole{\frac{1}{\ep^8}} \Big(
   \mfrac{1}{192}
 \Big)+ \pole{\frac{1}{\ep^6}} \Big(
  - \mfrac{19}{72}\zeta_2
 \Big)+ \pole{\frac{1}{\ep^5}} \Big(
  - \mfrac{61}{32}\zeta_3
 \Big)+ \pole{\frac{1}{\ep^4}} \Big(
  - \mfrac{5089}{1440}\zeta_2^2
 \Big)+ \pole{\frac{1}{\ep^3}} \Big(
  - \mfrac{41237}{480}\zeta_5
  + \mfrac{4111}{216}\zeta_3 \zeta_2
 \Big)  \notag\\ &\quad
+ \pole{\frac{1}{\ep^2}} \Big(
  - \mfrac{2881}{864}\zeta_3^2
  - \mfrac{8259}{112}\zeta_2^3
 \Big)+ \pole{\frac{1}{\ep}} \Big(
  - \mfrac{819241}{224}\zeta_7
  + \mfrac{314971}{360}\zeta_5 \zeta_2
  + \mfrac{325133}{2160}\zeta_3 \zeta_2^2
 \Big)+ \pole{} \Big(
   \mfrac{27142}{15}\zeta_{5,3}
  \notag\\ &\quad
  + \mfrac{1433501}{240}\zeta_5 \zeta_3
  + \mfrac{88663}{162}\zeta_3^2 \zeta_2
  - \mfrac{994787867}{504000}\zeta_2^4
 \Big) + \order{\ep},\\
I_{3}^{(23)} &=  \pole{\frac{1}{\ep^8}} \Big(
   \mfrac{1}{144}
 \Big)+ \pole{\frac{1}{\ep^6}} \Big(
  - \mfrac{5}{18}\zeta_2
 \Big)+ \pole{\frac{1}{\ep^5}} \Big(
  - \mfrac{401}{216}\zeta_3
 \Big)+ \pole{\frac{1}{\ep^4}} \Big(
   \mfrac{19}{16}\zeta_2^2
 \Big)+ \pole{\frac{1}{\ep^3}} \Big(
  - \mfrac{16277}{360}\zeta_5
  + \mfrac{13151}{216}\zeta_3 \zeta_2
 \Big)  \notag\\ &\quad
+ \pole{\frac{1}{\ep^2}} \Big(
   \mfrac{248513}{1296}\zeta_3^2
  + \mfrac{751}{45}\zeta_2^3
 \Big)+ \pole{\frac{1}{\ep}} \Big(
  - \mfrac{2796859}{4032}\zeta_7
  + \mfrac{37751}{36}\zeta_5 \zeta_2
  - \mfrac{653}{5}\zeta_3 \zeta_2^2
 \Big)+ \pole{} \Big(
  - \mfrac{39277}{60}\zeta_{5,3}
  + \mfrac{5465129}{1080}\zeta_5 \zeta_3
  \notag\\ &\quad
  - \mfrac{378593}{81}\zeta_3^2 \zeta_2
  + \mfrac{53058307}{126000}\zeta_2^4
 \Big) + \order{\ep},\\
I_{4}^{(24)} &=  \pole{\frac{1}{\ep^8}} \Big(
  - \mfrac{5}{576}
 \Big)+ \pole{\frac{1}{\ep^6}} \Big(
   \mfrac{65}{144}\zeta_2
 \Big)+ \pole{\frac{1}{\ep^5}} \Big(
   \mfrac{1645}{864}\zeta_3
 \Big)+ \pole{\frac{1}{\ep^4}} \Big(
  - \mfrac{109}{40}\zeta_2^2
 \Big)+ \pole{\frac{1}{\ep^3}} \Big(
   \mfrac{2093}{288}\zeta_5
  - \mfrac{9361}{216}\zeta_3 \zeta_2
 \Big)  \notag\\ &\quad
+ \pole{\frac{1}{\ep^2}} \Big(
  - \mfrac{166229}{2592}\zeta_3^2
  - \mfrac{289223}{10080}\zeta_2^3
 \Big)+ \pole{\frac{1}{\ep}} \Big(
   \mfrac{995315}{4032}\zeta_7
  - \mfrac{24133}{36}\zeta_5 \zeta_2
  + \mfrac{39527}{120}\zeta_3 \zeta_2^2
 \Big)+ \pole{} \Big(
  - \mfrac{1533}{40}\zeta_{5,3}
  \notag\\ &\quad
  - \mfrac{339469}{432}\zeta_5 \zeta_3
  + \mfrac{134365}{81}\zeta_3^2 \zeta_2
  + \mfrac{29437571}{1008000}\zeta_2^4
 \Big) + \order{\ep},\\
I_{5}^{(25)} &= I_{\mathrm{p}, 19}^{(25)},\\
I_{6}^{(26)} &=  \pole{\frac{1}{\ep^8}} \Big(
  - \mfrac{25}{576}
 \Big)+ \pole{\frac{1}{\ep^6}} \Big(
   \mfrac{313}{288}\zeta_2
 \Big)+ \pole{\frac{1}{\ep^5}} \Big(
   \mfrac{1241}{216}\zeta_3
 \Big)+ \pole{\frac{1}{\ep^4}} \Big(
  - \mfrac{3671}{720}\zeta_2^2
 \Big)+ \pole{\frac{1}{\ep^3}} \Big(
   \mfrac{275}{9}\zeta_5
  - \mfrac{7033}{54}\zeta_3 \zeta_2
 \Big)  \notag\\ &\quad
+ \pole{\frac{1}{\ep^2}} \Big(
  - \mfrac{210031}{648}\zeta_3^2
  - \mfrac{9349}{105}\zeta_2^3
 \Big)+ \pole{\frac{1}{\ep}} \Big(
   \mfrac{3509717}{2016}\zeta_7
  - \mfrac{366929}{180}\zeta_5 \zeta_2
  + \mfrac{284633}{540}\zeta_3 \zeta_2^2
 \Big)+ \pole{} \Big(
  - \mfrac{10763}{6}\zeta_{5,3}
  \notag\\ &\quad
  - \mfrac{3150517}{540}\zeta_5 \zeta_3
  + \mfrac{1847833}{324}\zeta_3^2 \zeta_2
  + \mfrac{2984111}{1575}\zeta_2^4
 \Big) + \order{\ep},\\
I_{7}^{(26)} &=  \pole{\frac{1}{\ep^8}} \Big(
   \mfrac{1}{288}
 \Big)+ \pole{\frac{1}{\ep^6}} \Big(
   \mfrac{1}{144}\zeta_2
 \Big)+ \pole{\frac{1}{\ep^5}} \Big(
   \mfrac{209}{216}\zeta_3
 \Big)+ \pole{\frac{1}{\ep^4}} \Big(
   \mfrac{43}{40}\zeta_2^2
 \Big)+ \pole{\frac{1}{\ep^3}} \Big(
  - \mfrac{5761}{360}\zeta_5
  + \mfrac{59}{27}\zeta_3 \zeta_2
 \Big)+ \pole{\frac{1}{\ep^2}} \Big(
   \mfrac{27179}{648}\zeta_3^2
  \notag\\ &\quad
  - \mfrac{17501}{2520}\zeta_2^3
 \Big)+ \pole{\frac{1}{\ep}} \Big(
   \mfrac{4704689}{2016}\zeta_7
  + \mfrac{22139}{180}\zeta_5 \zeta_2
  + \mfrac{11471}{60}\zeta_3 \zeta_2^2
 \Big)+ \pole{} \Big(
  - \mfrac{64211}{10}\zeta_{5,3}
  + \mfrac{4526447}{540}\zeta_5 \zeta_3
  - \mfrac{41461}{81}\zeta_3^2 \zeta_2
  \notag\\ &\quad
  + \mfrac{1215668297}{252000}\zeta_2^4
 \Big) + \order{\ep},\\
I_{8}^{(27)} &=  \pole{\frac{1}{\ep^8}} \Big(
  - \mfrac{1}{64}
 \Big)+ \pole{\frac{1}{\ep^6}} \Big(
   \mfrac{5}{24}\zeta_2
 \Big)+ \pole{\frac{1}{\ep^5}} \Big(
   \mfrac{55}{48}\zeta_3
 \Big)+ \pole{\frac{1}{\ep^4}} \Big(
   \mfrac{49}{320}\zeta_2^2
 \Big)+ \pole{\frac{1}{\ep^3}} \Big(
   \mfrac{1183}{240}\zeta_5
  - \mfrac{3395}{288}\zeta_3 \zeta_2
 \Big)+ \pole{\frac{1}{\ep^2}} \Big(
  - \mfrac{20561}{576}\zeta_3^2
  \notag\\ &\quad
  - \mfrac{39377}{1680}\zeta_2^3
 \Big)+ \pole{\frac{1}{\ep}} \Big(
  - \mfrac{8686441}{5376}\zeta_7
  + \mfrac{217}{6}\zeta_5 \zeta_2
  + \mfrac{1617}{20}\zeta_3 \zeta_2^2
 \Big)+ \pole{} \Big(
   \mfrac{233423}{80}\zeta_{5,3}
  + \mfrac{882581}{288}\zeta_5 \zeta_3
  + \mfrac{136097}{108}\zeta_3^2 \zeta_2
  \notag\\ &\quad
  - \mfrac{412195213}{168000}\zeta_2^4
 \Big) + \order{\ep},\\
I_{9}^{(28)} &=  \pole{\frac{1}{\ep^8}} \Big(
  - \mfrac{1}{96}
 \Big)+ \pole{\frac{1}{\ep^6}} \Big(
   \mfrac{97}{288}\zeta_2
 \Big)+ \pole{\frac{1}{\ep^5}} \Big(
   \mfrac{271}{144}\zeta_3
 \Big)+ \pole{\frac{1}{\ep^4}} \Big(
  - \mfrac{3793}{2880}\zeta_2^2
 \Big)+ \pole{\frac{1}{\ep^3}} \Big(
   \mfrac{4291}{120}\zeta_5
  - \mfrac{21359}{432}\zeta_3 \zeta_2
 \Big)  \notag\\ &\quad
+ \pole{\frac{1}{\ep^2}} \Big(
  - \mfrac{19235}{144}\zeta_3^2
  + \mfrac{1397}{576}\zeta_2^3
 \Big)+ \pole{\frac{1}{\ep}} \Big(
   \mfrac{909513}{896}\zeta_7
  - \mfrac{420149}{720}\zeta_5 \zeta_2
  + \mfrac{487043}{4320}\zeta_3 \zeta_2^2
 \Big)+ \pole{} \Big(
  - \mfrac{23879}{40}\zeta_{5,3}
  \notag\\ &\quad
  - \mfrac{1003741}{180}\zeta_5 \zeta_3
  + \mfrac{4318261}{1296}\zeta_3^2 \zeta_2
  + \mfrac{1298874221}{2016000}\zeta_2^4
 \Big) + \order{\ep},\\
I_{10}^{(29)} &=  \pole{\frac{1}{\ep^8}} \Big(
  - \mfrac{1}{1152}
 \Big)+ \pole{\frac{1}{\ep^6}} \Big(
  - \mfrac{1}{576}\zeta_2
 \Big)+ \pole{\frac{1}{\ep^5}} \Big(
  - \mfrac{13}{1728}\zeta_3
 \Big)+ \pole{\frac{1}{\ep^4}} \Big(
   \mfrac{169}{640}\zeta_2^2
 \Big)+ \pole{\frac{1}{\ep^3}} \Big(
   \mfrac{26357}{2880}\zeta_5
  + \mfrac{685}{1728}\zeta_3 \zeta_2
 \Big)  \notag\\ &\quad
+ \pole{\frac{1}{\ep^2}} \Big(
   \mfrac{186637}{10368}\zeta_3^2
  + \mfrac{57191}{5040}\zeta_2^3
 \Big)+ \pole{\frac{1}{\ep}} \Big(
   \mfrac{4990045}{8064}\zeta_7
  - \mfrac{504581}{2880}\zeta_5 \zeta_2
  + \mfrac{13379}{960}\zeta_3 \zeta_2^2
 \Big)+ \pole{} \Big(
  - \mfrac{86383}{480}\zeta_{5,3}
  \notag\\ &\quad
  + \mfrac{1612597}{8640}\zeta_5 \zeta_3
  - \mfrac{1324753}{2592}\zeta_3^2 \zeta_2
  + \mfrac{134849039}{504000}\zeta_2^4
 \Big) + \order{\ep},\\
I_{11}^{(30)} &= I_{\mathrm{p}, 20}^{(30)},\\
I_{12}^{(27)} &=  \pole{\frac{1}{\ep^8}} \Big(
   \mfrac{35}{1152}
 \Big)+ \pole{\frac{1}{\ep^6}} \Big(
  - \mfrac{73}{192}\zeta_2
 \Big)+ \pole{\frac{1}{\ep^5}} \Big(
  - \mfrac{1015}{432}\zeta_3
 \Big)+ \pole{\frac{1}{\ep^4}} \Big(
  - \mfrac{4069}{1440}\zeta_2^2
 \Big)+ \pole{\frac{1}{\ep^3}} \Big(
  - \mfrac{8693}{144}\zeta_5
  + \mfrac{5809}{288}\zeta_3 \zeta_2
 \Big)  \notag\\ &\quad
+ \pole{\frac{1}{\ep^2}} \Big(
   \mfrac{260783}{5184}\zeta_3^2
  - \mfrac{36499}{2240}\zeta_2^3
 \Big)+ \pole{\frac{1}{\ep}} \Big(
  - \mfrac{57455}{288}\zeta_7
  + \mfrac{255661}{480}\zeta_5 \zeta_2
  + \mfrac{85981}{432}\zeta_3 \zeta_2^2
 \Big)+ \pole{} \Big(
  - \mfrac{10607}{5}\zeta_{5,3}
  \notag\\ &\quad
  + \mfrac{2369983}{864}\zeta_5 \zeta_3
  - \mfrac{23653}{48}\zeta_3^2 \zeta_2
  + \mfrac{4065832699}{2016000}\zeta_2^4
 \Big) + \order{\ep},\\
I_{13}^{(28)} &=  \pole{\frac{1}{\ep^8}} \Big(
  - \mfrac{13}{1152}
 \Big)+ \pole{\frac{1}{\ep^6}} \Big(
   \mfrac{35}{192}\zeta_2
 \Big)+ \pole{\frac{1}{\ep^5}} \Big(
   \mfrac{305}{432}\zeta_3
 \Big)+ \pole{\frac{1}{\ep^4}} \Big(
   \mfrac{461}{1440}\zeta_2^2
 \Big)+ \pole{\frac{1}{\ep^3}} \Big(
   \mfrac{4001}{180}\zeta_5
  - \mfrac{461}{72}\zeta_3 \zeta_2
 \Big)+ \pole{\frac{1}{\ep^2}} \Big(
   \mfrac{11243}{324}\zeta_3^2
  \notag\\ &\quad
  + \mfrac{4295}{336}\zeta_2^3
 \Big)+ \pole{\frac{1}{\ep}} \Big(
   \mfrac{1798807}{8064}\zeta_7
  - \mfrac{3411}{16}\zeta_5 \zeta_2
  + \mfrac{2686}{27}\zeta_3 \zeta_2^2
 \Big)+ \pole{} \Big(
   \mfrac{535}{8}\zeta_{5,3}
  - \mfrac{229363}{108}\zeta_5 \zeta_3
  - \mfrac{34495}{144}\zeta_3^2 \zeta_2
  \notag\\ &\quad
  - \mfrac{12987991}{50400}\zeta_2^4
 \Big) + \order{\ep},\\
I_{14}^{(29)} &=  \pole{\frac{1}{\ep^5}} \Big(
  - \mfrac{1}{32}\zeta_3
 \Big)+ \pole{\frac{1}{\ep^4}} \Big(
   \mfrac{9}{320}\zeta_2^2
 \Big)+ \pole{\frac{1}{\ep^3}} \Big(
  - \mfrac{371}{96}\zeta_5
  + \mfrac{91}{48}\zeta_3 \zeta_2
 \Big)+ \pole{\frac{1}{\ep^2}} \Big(
  - \mfrac{223}{96}\zeta_3^2
  + \mfrac{653}{576}\zeta_2^3
 \Big)+ \pole{\frac{1}{\ep}} \Big(
  - \mfrac{4871}{48}\zeta_7
  \notag\\ &\quad
  + \mfrac{789}{8}\zeta_5 \zeta_2
  + \mfrac{2287}{480}\zeta_3 \zeta_2^2
 \Big)+ \pole{} \Big(
   \mfrac{7507}{60}\zeta_{5,3}
  + \mfrac{421943}{720}\zeta_5 \zeta_3
  + \mfrac{2945}{144}\zeta_3^2 \zeta_2
  + \mfrac{13457111}{672000}\zeta_2^4
 \Big) + \order{\ep},\\
I_{15}^{(29)} &=  \pole{\frac{1}{\ep^8}} \Big(
  - \mfrac{1}{18}
 \Big)+ \pole{\frac{1}{\ep^6}} \Big(
   \mfrac{53}{48}\zeta_2
 \Big)+ \pole{\frac{1}{\ep^5}} \Big(
   \mfrac{2621}{432}\zeta_3
 \Big)+ \pole{\frac{1}{\ep^4}} \Big(
  - \mfrac{1423}{1440}\zeta_2^2
 \Big)+ \pole{\frac{1}{\ep^3}} \Big(
   \mfrac{54437}{720}\zeta_5
  - \mfrac{7751}{72}\zeta_3 \zeta_2
 \Big)  \notag\\ &\quad
+ \pole{\frac{1}{\ep^2}} \Big(
  - \mfrac{413683}{1296}\zeta_3^2
  - \mfrac{410153}{10080}\zeta_2^3
 \Big)+ \pole{\frac{1}{\ep}} \Big(
   \mfrac{12394561}{8064}\zeta_7
  - \mfrac{386357}{240}\zeta_5 \zeta_2
  - \mfrac{221929}{2160}\zeta_3 \zeta_2^2
 \Big)+ \pole{} \Big(
   \mfrac{13907}{120}\zeta_{5,3}
  \notag\\ &\quad
  - \mfrac{1866583}{270}\zeta_5 \zeta_3
  + \mfrac{664117}{144}\zeta_3^2 \zeta_2
  - \mfrac{251120053}{1008000}\zeta_2^4
 \Big) + \order{\ep},\\
I_{16}^{(30)} &=  \pole{\frac{1}{\ep^8}} \Big(
   \mfrac{7}{192}
 \Big)+ \pole{\frac{1}{\ep^6}} \Big(
  - \mfrac{35}{96}\zeta_2
 \Big)+ \pole{\frac{1}{\ep^5}} \Big(
  - \mfrac{271}{144}\zeta_3
 \Big)+ \pole{\frac{1}{\ep^4}} \Big(
  - \mfrac{49}{160}\zeta_2^2
 \Big)+ \pole{\frac{1}{\ep^3}} \Big(
  - \mfrac{6037}{240}\zeta_5
  + \mfrac{3343}{144}\zeta_3 \zeta_2
 \Big)  \notag\\ &\quad
+ \pole{\frac{1}{\ep^2}} \Big(
   \mfrac{42271}{864}\zeta_3^2
  + \mfrac{1711}{315}\zeta_2^3
 \Big)+ \pole{\frac{1}{\ep}} \Big(
  - \mfrac{32891}{96}\zeta_7
  + \mfrac{22439}{48}\zeta_5 \zeta_2
  - \mfrac{3241}{48}\zeta_3 \zeta_2^2
 \Big)+ \pole{} \Big(
   \mint{-256}\zeta_{5,3}
  + \mfrac{280219}{720}\zeta_5 \zeta_3
  \notag\\ &\quad
  - \mfrac{166087}{216}\zeta_3^2 \zeta_2
  + \mfrac{2656651}{16800}\zeta_2^4
 \Big) + \order{\ep},\\
I_{17}^{(30)} &=  \pole{\frac{1}{\ep^5}} \Big(
  - \mfrac{1}{32}\zeta_3
 \Big)+ \pole{\frac{1}{\ep^4}} \Big(
   \mfrac{37}{960}\zeta_2^2
 \Big)+ \pole{\frac{1}{\ep^3}} \Big(
   \mfrac{49}{96}\zeta_5
  + \mfrac{9}{16}\zeta_3 \zeta_2
 \Big)+ \pole{\frac{1}{\ep^2}} \Big(
  - \mfrac{625}{96}\zeta_3^2
  + \mfrac{81401}{20160}\zeta_2^3
 \Big)+ \pole{\frac{1}{\ep}} \Big(
   \mfrac{307}{16}\zeta_7
  + \mfrac{703}{6}\zeta_5 \zeta_2
  \notag\\ &\quad
  - \mfrac{81341}{1440}\zeta_3 \zeta_2^2
 \Big)+ \pole{} \Big(
  - \mfrac{3505}{24}\zeta_{5,3}
  - \mfrac{505507}{720}\zeta_5 \zeta_3
  + \mfrac{3259}{48}\zeta_3^2 \zeta_2
  + \mfrac{57616759}{403200}\zeta_2^4
 \Big) + \order{\ep},\\
I_{18}^{(30)} &=  \pole{\frac{1}{\ep^8}} \Big(
   \mfrac{1}{288}
 \Big)+ \pole{\frac{1}{\ep^6}} \Big(
   \mfrac{11}{288}\zeta_2
 \Big)+ \pole{\frac{1}{\ep^5}} \Big(
  - \mfrac{1}{864}\zeta_3
 \Big)+ \pole{\frac{1}{\ep^4}} \Big(
  - \mfrac{241}{160}\zeta_2^2
 \Big)+ \pole{\frac{1}{\ep^3}} \Big(
  - \mfrac{18559}{1440}\zeta_5
  - \mfrac{9749}{864}\zeta_3 \zeta_2
 \Big)  \notag\\ &\quad
+ \pole{\frac{1}{\ep^2}} \Big(
  - \mfrac{153467}{5184}\zeta_3^2
  - \mfrac{763019}{20160}\zeta_2^3
 \Big)+ \pole{\frac{1}{\ep}} \Big(
  - \mfrac{1102943}{1008}\zeta_7
  - \mfrac{249299}{1440}\zeta_5 \zeta_2
  + \mfrac{2177}{30}\zeta_3 \zeta_2^2
 \Big)+ \pole{} \Big(
   \mfrac{6098}{15}\zeta_{5,3}
  \notag\\ &\quad
  - \mfrac{5763491}{4320}\zeta_5 \zeta_3
  + \mfrac{1405787}{1296}\zeta_3^2 \zeta_2
  - \mfrac{2039554577}{2016000}\zeta_2^4
 \Big) + \order{\ep},\\
I_{19}^{(22)} &=  \pole{\frac{1}{\ep^8}} \Big(
   \mfrac{1}{576}
 \Big)+ \pole{\frac{1}{\ep^6}} \Big(
   \mfrac{29}{288}\zeta_2
 \Big)+ \pole{\frac{1}{\ep^5}} \Big(
   \mfrac{269}{432}\zeta_3
 \Big)+ \pole{\frac{1}{\ep^4}} \Big(
   \mfrac{553}{360}\zeta_2^2
 \Big)+ \pole{\frac{1}{\ep^3}} \Big(
   \mfrac{43109}{720}\zeta_5
  - \mfrac{349}{27}\zeta_3 \zeta_2
 \Big)+ \pole{\frac{1}{\ep^2}} \Big(
   \mfrac{57485}{1296}\zeta_3^2
  \notag\\ &\quad
  + \mfrac{142267}{2520}\zeta_2^3
 \Big)+ \pole{\frac{1}{\ep}} \Big(
   \mfrac{995699}{252}\zeta_7
  - \mfrac{49408}{45}\zeta_5 \zeta_2
  + \mfrac{2279}{108}\zeta_3 \zeta_2^2
 \Big)+ \pole{} \Big(
  - \mfrac{17607}{5}\zeta_{5,3}
  - \mfrac{6097903}{1080}\zeta_5 \zeta_3
  - \mfrac{980735}{648}\zeta_3^2 \zeta_2
  \notag\\ &\quad
  + \mfrac{203235527}{63000}\zeta_2^4
 \Big) + \order{\ep},\\
I_{20}^{(22)} &=  \pole{\frac{1}{\ep^2}} \Big(
   \mfrac{1}{4}\zeta_3^2
  + \mfrac{31}{140}\zeta_2^3
 \Big)+ \pole{\frac{1}{\ep}} \Big(
  - \mfrac{199}{64}\zeta_7
  - \mfrac{39}{8}\zeta_5 \zeta_2
  + \mfrac{7}{5}\zeta_3 \zeta_2^2
 \Big)+ \pole{} \Big(
   \mfrac{78}{5}\zeta_{5,3}
  + \mint{-61}\zeta_5 \zeta_3
  + \mfrac{13}{8}\zeta_3^2 \zeta_2
  - \mfrac{515773}{28000}\zeta_2^4
 \Big)  \notag\\ &\quad
 + \order{\ep},\\
I_{21}^{(24)} &=  \pole{\frac{1}{\ep^8}} \Big(
   \mfrac{5}{576}
 \Big)+ \pole{\frac{1}{\ep^6}} \Big(
   \mfrac{37}{288}\zeta_2
 \Big)+ \pole{\frac{1}{\ep^5}} \Big(
   \mfrac{229}{432}\zeta_3
 \Big)+ \pole{\frac{1}{\ep^4}} \Big(
  - \mfrac{541}{360}\zeta_2^2
 \Big)+ \pole{\frac{1}{\ep^3}} \Big(
   \mfrac{1799}{144}\zeta_5
  - \mfrac{13385}{432}\zeta_3 \zeta_2
 \Big)+ \pole{\frac{1}{\ep^2}} \Big(
  - \mfrac{259405}{2592}\zeta_3^2
  \notag\\ &\quad
  - \mfrac{222371}{10080}\zeta_2^3
 \Big)+ \pole{\frac{1}{\ep}} \Big(
   \mfrac{858061}{4032}\zeta_7
  - \mfrac{397679}{720}\zeta_5 \zeta_2
  - \mfrac{61453}{2160}\zeta_3 \zeta_2^2
 \Big)+ \pole{} \Big(
   \mfrac{7949}{24}\zeta_{5,3}
  - \mfrac{5618761}{2160}\zeta_5 \zeta_3
  + \mfrac{2322067}{1296}\zeta_3^2 \zeta_2
  \notag\\ &\quad
  - \mfrac{27047381}{67200}\zeta_2^4
 \Big) + \order{\ep},\\
I_{22}^{(24)} &=  \pole{\frac{1}{\ep^8}} \Big(
   \mfrac{1}{144}
 \Big)+ \pole{\frac{1}{\ep^6}} \Big(
  - \mfrac{1}{18}\zeta_2
 \Big)+ \pole{\frac{1}{\ep^5}} \Big(
  - \mfrac{263}{216}\zeta_3
 \Big)+ \pole{\frac{1}{\ep^4}} \Big(
  - \mfrac{3127}{720}\zeta_2^2
 \Big)+ \pole{\frac{1}{\ep^3}} \Big(
  - \mfrac{27287}{360}\zeta_5
  + \mfrac{683}{108}\zeta_3 \zeta_2
 \Big)  \notag\\ &\quad
+ \pole{\frac{1}{\ep^2}} \Big(
   \mfrac{35743}{648}\zeta_3^2
  - \mfrac{71705}{1008}\zeta_2^3
 \Big)+ \pole{\frac{1}{\ep}} \Big(
  - \mfrac{9118279}{4032}\zeta_7
  + \mfrac{161671}{360}\zeta_5 \zeta_2
  + \mfrac{483137}{1080}\zeta_3 \zeta_2^2
 \Big)+ \pole{} \Big(
   \mfrac{2023}{12}\zeta_{5,3}
  + \mfrac{4183531}{540}\zeta_5 \zeta_3
  \notag\\ &\quad
  + \mfrac{109561}{648}\zeta_3^2 \zeta_2
  - \mfrac{30490193}{100800}\zeta_2^4
 \Big) + \order{\ep},\\
I_{23}^{(28)} &=  \pole{\frac{1}{\ep^4}} \Big(
   \mfrac{1}{4}\zeta_2^2
 \Big)+ \pole{\frac{1}{\ep^3}} \Big(
   \mfrac{5}{4}\zeta_5
  + \mfrac{3}{4}\zeta_3 \zeta_2
 \Big)+ \pole{\frac{1}{\ep^2}} \Big(
   \mfrac{117}{8}\zeta_3^2
  - \mfrac{659}{168}\zeta_2^3
 \Big)+ \pole{\frac{1}{\ep}} \Big(
  - \mint{65}\zeta_7
  - \mfrac{443}{4}\zeta_5 \zeta_2
  + \mfrac{421}{6}\zeta_3 \zeta_2^2
 \Big)+ \pole{} \Big(
   \mint{69}\zeta_{5,3}
  \notag\\ &\quad
  + \mfrac{9349}{12}\zeta_5 \zeta_3
  - \mfrac{111}{2}\zeta_3^2 \zeta_2
  - \mfrac{780079}{5600}\zeta_2^4
 \Big) + \order{\ep}.

\label{eq:intsollast}
\end{align}
}
We also provide these analytical solutions in the ancillary files for this paper.
Our results are expressed in terms of regular and multiple zeta values
\begin{align}
\zeta_i &= \sum_{m=1}^\infty \frac{1}{m^i}~~(i=2,\ldots,7), &
    \zeta_{5,3} &= \sum_{m=1}^\infty \sum_{n=1}^{m-1} \frac{1}{m^5 n^3}
    \approx 0.0377076729848\,. 
\end{align}
We see from Eqs.~\eqref{eq:intsolfirst}-\eqref{eq:intsollast} that the integrals $I^{(n_i)}_i$ through to weight 8 are of uniform transcendental weight.
Moreover, we reduced several of the master integrals $I^{(n_i)}_i$ to the UT basis integrals employed in our differential equations calculations and found the coefficients to be rational numbers, showing that their UT property extends to higher orders in $\epsilon$ as well.

\section{Result for the Sudakov form factor}
\label{sec:results}

Inserting our results for the master integrals in the reduced expression~\eqref{eq:integrand} 
we obtain for the four-loop contribution to the Sudakov form factor
\begin{align}
\label{eq:ff4lres}
F_4 &= \Big[
 \pole{\frac{1}{\ep^8}} \Big(
   \mfrac{2}{3}
 \Big)+ \pole{\frac{1}{\ep^6}} \Big(
   \mfrac{2}{3}\zeta_2
 \Big)+ \pole{\frac{1}{\ep^5}} \Big(
  - \mfrac{38}{9}\zeta_3
 \Big)+ \pole{\frac{1}{\ep^4}} \Big(
   \mfrac{5}{18}\zeta_2^2
 \Big)+ \pole{\frac{1}{\ep^3}} \Big(
   \mfrac{1082}{15}\zeta_5
  + \mfrac{23}{3}\zeta_3 \zeta_2
 \Big)
\notag\\ &\quad 
 + \pole{\frac{1}{\ep^2}} \Big(
   \mfrac{10853}{54}\zeta_3^2
  + \mfrac{95477}{945}\zeta_2^3
 \Big) + \pole{\frac{1}{\ep}} \Big(
   \mfrac{541619}{126}\zeta_7
  - \mfrac{15529}{45}\zeta_5 \zeta_2
  + \mfrac{39067}{135}\zeta_3 \zeta_2^2
 \Big)+ \pole{} \Big(
  - \mfrac{808}{45}\zeta_{5,3}
\notag\\ &\quad
  + \mfrac{499927}{45}\zeta_5 \zeta_3
  - \mfrac{35707}{27}\zeta_3^2 \zeta_2
  + \mfrac{71888861}{31500}\zeta_2^4
 \Big) \Big]
 + \casimir{\frac{1}{N_c^2}}\Big[
  \pole{\frac{1}{\ep^2}} \Big(
   \mint{18}\zeta_3^2
  + \mfrac{372}{35}\zeta_2^3
 \Big)+ \pole{\frac{1}{\ep}} \Big(
  - \mfrac{2613}{4}\zeta_7
\notag\\ &\quad
  - \mint{192}\zeta_5 \zeta_2
  + \mfrac{138}{5}\zeta_3 \zeta_2^2
 \Big)
 + \pole{} \Big(
   \mint{390}\zeta_{5,3}
  - \mint{7638}\zeta_5 \zeta_3
  - \mint{24}\zeta_3^2 \zeta_2
  - \mfrac{248383}{175}\zeta_2^4
 \Big) + \order{\ep}
 \Big].
\end{align}
While the poles in $\epsilon$ are known, the finite term of the form factor is new and represents the main result of this paper.

It is useful to consider the logarithm of the form factor
\begin{align}
\ln(F) &= \ln\left(1 + \pole{a}z^{\ep} F_1  + \pole{a^2}z^{2\ep} F_2  + \pole{a^3}z^{3\ep} F_3  + \pole{a^4}z^{4\ep} F_4 + \ldots \right)\\
&= \pole{a} z^\ep F_1
 + \pole{a^2}z^{2\ep} \big( F_2 - \mfrac{1}{2} F_1^2 \big)
 + \pole{a^3}z^{3\ep} \big( F_3 - F_2 F_1 + \mfrac{1}{3} F_1^3 \big)
 + \pole{a^4}z^{4\ep} \big( F_4 - F_3 F_1 - \mfrac{1}{2}F_2^2 \notag\\ &\quad
 + F_2 F_1^2  - \mfrac{1}{4} F_1^4 \big) + \ldots,
 \end{align}
which can be written in the form~\cite{Moch:2005id,Ravindran:2006cg}
\begin{align}
\ln(F) &= \sum_{L=1}^\infty \pole{a^L} z^{L\ep} \left(
 - \frac{\Gamma_L}{2 (L \pole{\ep})^2} - \frac{\mathcal{G}_L}{ 2 L \pole{\ep} }
 + L^\text{fin}_L \right),
\end{align}
defining the finite remainders $L^\text{fin}_L$.
The pole terms are fixed by the cusp anomalous dimension~\cite{Henn:2019swt,Huber:2019fxe}
\begin{align}
\Gamma &= \sum_{L=1}^\infty \pole{a^L} \Gamma_L \\
  &= \pole{a} \Big(\mint{4}\Big)+ \pole{a^2} \Big(\mint{-8}\zeta_2\Big)
  + \pole{a^3} \Big(\mint{88}\zeta_4 \Big)
  + \pole{a^4} \Big(- \mint{876} \zeta_6 - \mint{32}\zeta_3^2
  + \frac{1}{N_c^2} \Big[ - \mint{1488} \zeta_6 - \mint{576}\zeta_3^2 \Big]\Big)
\notag\\ &\quad  + \order{a^5}
\end{align}
and the collinear anomalous dimension~\cite{Agarwal:2021zft}
\begin{align}
\mathcal{G} &= \sum_{L=1}^\infty \pole{a^L} \mathcal{G}_L \\
&= \pole{a^2} \Big(- \mint{4}\zeta_3\Big)+ \pole{a^3} \Big(\mint{32}\zeta_5
 + \mfrac{80}{3}\zeta_3 \zeta_2 \Big)+ \pole{a^4} \Big(
  - \mint{300}\zeta_7 - \mint{256}\zeta_5 \zeta_2  - \mint{384} \zeta_4 \zeta_3
  + \frac{1}{N_c^2} \Big[ \mint{5226}\zeta_7 \notag\\ &\quad
  + \mint{1536}\zeta_5\zeta_2 -\mint{552}\zeta_4\zeta_3 \Big]\Big) +\order{a^5} \,.
\end{align}
Using the one-, two- and three-loop results of~\cite{Gehrmann:2011xn} and our four-loop result~\eqref{eq:ff4lres}, we find for the finite remainders of the logarithm of the form factor
\begin{align}
L^\text{fin}_1 &= \Big(
   \zeta_2
 \Big)+ \pole{\ep} \Big(
   \mfrac{14}{3}\zeta_3
 \Big)+ \pole{\ep^2} \Big(
   \mfrac{47}{20}\zeta_2^2
 \Big)+ \pole{\ep^3} \Big(
   \mfrac{62}{5}\zeta_5
  - \mfrac{7}{3}\zeta_3 \zeta_2
 \Big)+ \pole{\ep^4} \Big(
  - \mfrac{49}{9}\zeta_3^2
  + \mfrac{949}{280}\zeta_2^3
 \Big)+ \pole{\ep^5} \Big(
   \mfrac{254}{7}\zeta_7
  \notag\\ &\quad
  - \mfrac{31}{5}\zeta_5 \zeta_2
  - \mfrac{329}{60}\zeta_3 \zeta_2^2
 \Big)+ \pole{\ep^6} \Big(
  - \mfrac{434}{15}\zeta_5 \zeta_3
  + \mfrac{49}{18}\zeta_3^2 \zeta_2
  + \mfrac{55779}{11200}\zeta_2^4
 \Big) + \order{\ep^7}\,,
 \\
L^\text{fin}_2 &=
\pole{\ep} \Big(
   \mint{39}\zeta_5
  - \mfrac{5}{3}\zeta_3 \zeta_2
 \Big)+ \pole{\ep^2} \Big(
   \mfrac{235}{3}\zeta_3^2
  + \mfrac{2623}{70}\zeta_2^3
 \Big)+ \pole{\ep^3} \Big(
   \mint{978}\zeta_7
  - \mfrac{437}{5}\zeta_5 \zeta_2
  + \mfrac{219}{2}\zeta_3 \zeta_2^2
 \Big)  \notag\\ &\quad
+ \pole{\ep^4} \Big(
   \mint{-264}\zeta_{5,3}
  + \mfrac{2238}{5}\zeta_5 \zeta_3
  - \mfrac{1351}{9}\zeta_3^2 \zeta_2
  + \mint{508}\zeta_2^4
 \Big) + \order{\ep^5}\,,
\\
L^\text{fin}_3 &= \Big(
  - \mfrac{104}{9}\zeta_3^2
  - \mfrac{12352}{945}\zeta_2^3
 \Big)+ \pole{\ep} \Big(
  - \mfrac{21181}{18}\zeta_7
  + \mfrac{748}{9}\zeta_5 \zeta_2
  - \mfrac{856}{45}\zeta_3 \zeta_2^2
 \Big)+ \pole{\ep^2} \Big(
  - \mfrac{11368}{45}\zeta_{5,3}
  \notag\\ &\quad
  - \mfrac{15376}{3}\zeta_5 \zeta_3
  + \mfrac{4228}{9}\zeta_3^2 \zeta_2
  - \mfrac{1989614}{3375}\zeta_2^4
 \Big) + \order{\ep^3}\,,
\\
\label{eq:fin4l}
L^\text{fin}_4 &= 
   \mfrac{24}{5}\zeta_{5,3}
  + \mint{184}\zeta_5 \zeta_3
  + \mint{41}\zeta_3^2 \zeta_2
  + \mfrac{1505381}{10500}\zeta_2^4
+\frac{1}{N_c^2} \Big[
   \mint{390}\zeta_{5,3}
  - \mint{7638}\zeta_5 \zeta_3
  - \mint{24}\zeta_3^2 \zeta_2
  - \mfrac{248383}{175}\zeta_2^4
 \Big]
  \notag\\ &\quad
 + \order{\ep}\,.

\end{align}
We observe that the subtraction of exponentiated terms leads to somewhat simpler rational prefactors in the finite remainder~\eqref{eq:fin4l} compared to the weight 8 terms of~\eqref{eq:ff4lres}.

\section{Conclusions}
\label{sec:conclusions}

We presented the analytical calculation of the Sudakov form factor in $\mathcal{N}=4$ supersymmetric Yang-Mills theory to four loop order.
To solve the master integrals to weight 8, we employed direct parametric integrations and the method of differential equations with an auxiliary scale.

To the best of our knowledge, this is the first time that a form factor has been computed to four-loop order in full-color Yang-Mills theory, and we hope that our explicit results are helpful in further studies of formal and phenomenological theories at high perturbative orders.
The master integrals entering the present calculation form a subset of the most complicated integrals needed for general massless three-point functions with one off-shell leg.
Our methods allow us to calculate also the remaining master integrals, providing the last missing building block for the calculation of the massless corrections to the quark-photon vertex and the effective gluon-Higgs vertex in four-loop Quantum Chromodynamics.

\section*{Acknowledgments}

AvM and RMS gratefully acknowledge Erik Panzer for related collaborations.
This research was supported by the Deutsche Forschungsgemeinschaft (DFG,
German Research Foundation) under Grant No.\ 396021762 — TRR 257 ``Particle Physics
Phenomenology after the Higgs Discovery'' and by the National Science Foundation (NSF) under Grant No.\ 2013859 ``Multi-loop amplitudes and precise predictions for the LHC''.
The work of AVS and VAS was supported by the Russian Science Foundation under Agreement No.\ 21-71-30003
(IBP reduction) and by the Ministry of Education and Science of
the Russian Federation as part of the program of the Moscow Center for Fundamental and
Applied Mathematics under Agreement No.\ 075-15-2019-1621 (numerical checks of results for the master integrals with {\tt FIESTA}).
We acknowledge the High Performance Computing Center at Michigan State University
for computing resources.
The Feynman diagrams were drawn with the help of {\tt Axodraw}~\cite{Vermaseren:1994je}
and {\tt JaxoDraw}~\cite{Binosi:2003yf}.

\bibliographystyle{JHEP}
\bibliography{ff4lneq4fin}

\end{document}